\documentstyle[twoside]{article}

\catcode`\@=11
\long\def\@makefntext#1{
\protect\noindent \hbox to 3.2pt {\hskip-.9pt
$^{{\eightrm\@thefnmark}}$\hfil}#1\hfill}               

\def\@makefnmark{\hbox to 0pt{$^{\@thefnmark}$\hss}}    

\def\ps@myheadings{\let\@mkboth\@gobbletwo
\def\@oddhead{\hbox{}
\rightmark\hfil\eightrm\thepage}
\def\@oddfoot{}\def\@evenhead{\eightrm\thepage\hfil
\leftmark\hbox{}}\def\@evenfoot{}
\def\sectionmark##1{}\def\subsectionmark##1{}}

\oddsidemargin=\evensidemargin
\newcounter{sectionc}\newcounter{subsectionc}\newcounter{subsubsectionc}
\renewcommand{\section}[1] {\vspace{12pt}\addtocounter{sectionc}{1}
\setcounter{subsectionc}{0}\setcounter{subsubsectionc}{0}\noindent
        {\tenbf\thesectionc. #1}\par\vspace{5pt}}
\renewcommand{\subsection}[1] {\vspace{12pt}\addtocounter{subsectionc}{1}
        \setcounter{subsubsectionc}{0}\noindent
        {\bf\thesectionc.\thesubsectionc. {\kern1pt \bfit #1}}\par\vspace{5pt}}
\renewcommand{\subsubsection}[1] {\vspace{12pt}\addtocounter{subsubsectionc}{1}
        \noindent{\tenrm\thesectionc.\thesubsectionc.\thesubsubsectionc.
        {\kern1pt \tenit #1}}\par\vspace{5pt}}
\newcommand{\nonumsection}[1] {\vspace{12pt}\noindent{\tenbf #1}
        \par\vspace{5pt}}

\newcounter{appendixc}
\newcounter{subappendixc}[appendixc]
\newcounter{subsubappendixc}[subappendixc]
\renewcommand{\thesubappendixc}{\Alph{appendixc}.\arabic{subappendixc}}
\renewcommand{\thesubsubappendixc}
        {\Alph{appendixc}.\arabic{subappendixc}.\arabic{subsubappendixc}}

\renewcommand{\appendix}[1] {\vspace{12pt}
        \refstepcounter{appendixc}
        \setcounter{figure}{0}
        \setcounter{table}{0}
        \setcounter{lemma}{0}
        \setcounter{theorem}{0}
        \setcounter{corollary}{0}
        \setcounter{definition}{0}
        \setcounter{equation}{0}
        \renewcommand{\thefigure}{\Alph{appendixc}.\arabic{figure}}
        \renewcommand{\thetable}{\Alph{appendixc}.\arabic{table}}
        \renewcommand{\theappendixc}{\Alph{appendixc}}
        \renewcommand{\thelemma}{\Alph{appendixc}.\arabic{lemma}}
        \renewcommand{\thetheorem}{\Alph{appendixc}.\arabic{theorem}}
        \renewcommand{\thedefinition}{\Alph{appendixc}.\arabic{definition}}
        \renewcommand{\thecorollary}{\Alph{appendixc}.\arabic{corollary}}
        \renewcommand{\theequation}{\Alph{appendixc}.\arabic{equation}}
        \noindent{\tenbf Appendix \theappendixc #1}\par\vspace{5pt}}
\newcommand{\subappendix}[1] {\vspace{12pt}
        \refstepcounter{subappendixc}
        \noindent{\bf Appendix \thesubappendixc. {\kern1pt \bfit #1}}
        \par\vspace{5pt}}
\newcommand{\subsubappendix}[1] {\vspace{12pt}
        \refstepcounter{subsubappendixc}
        \noindent{\rm Appendix \thesubsubappendixc. {\kern1pt \tenit #1}}
        \par\vspace{5pt}}

\newcommand{\textlineskip}{\baselineskip=13pt}
\newcommand{\smalllineskip}{\baselineskip=10pt}

\def\eightcirc{
\begin{picture}(0,0)
\put(4.4,1.8){\circle{6.5}}
\end{picture}}
\def\eightcopyright{\eightcirc\kern2.7pt\hbox{\eightrm c}}

\newcommand{\copyrightheading}[1]
        {\vspace*{-2.5cm}\smalllineskip{\flushleft
        {\footnotesize International Journal of Modern Physics B, #1}\\
        {\footnotesize $\eightcopyright$\, World Scientific Publishing
         Company}\\
         }}

\newcommand{\publisher}[2]{{\begin{center}\footnotesize\smalllineskip
        Received #1\\
        Revised #2
        \end{center}
        }}

\def\abstracts#1#2#3{{
        \centering{\begin{minipage}{4.5in}\baselineskip=10pt\footnotesize
        \parindent=0pt #1\par
        \parindent=15pt #2\par
        \parindent=15pt #3
        \end{minipage}}\par}}

\renewenvironment{thebibliography}[1]                   
        {\frenchspacing
         \ninerm\baselineskip=11pt
         \begin{list}{\arabic{enumi}.}
        {\usecounter{enumi}\setlength{\parsep}{0pt}
         \setlength{\leftmargin 12.7pt}{\rightmargin 0pt} 
         \setlength{\itemsep}{0pt} \settowidth
        {\labelwidth}{#1.}\sloppy}}{\end{list}}

\newcounter{itemlistc}
\newcounter{romanlistc}
\newcounter{alphlistc}
\newcounter{arabiclistc}

\newcommand{\fcaption}[1]{
        \refstepcounter{figure}
        \setbox\@tempboxa = \hbox{\footnotesize Fig.~\thefigure. #1}
        \ifdim \wd\@tempboxa > 5in
           {\begin{center}
        \parbox{5in}{\footnotesize\smalllineskip Fig.~\thefigure. #1}
            \end{center}}
        \else
             {\begin{center}
             {\footnotesize Fig.~\thefigure. #1}
              \end{center}}
        \fi}

\newcommand{\tcaption}[1]{
        \refstepcounter{table}
        \setbox\@tempboxa = \hbox{\footnotesize Table~\thetable. #1}
        \ifdim \wd\@tempboxa > 5in
           {\begin{center}
        \parbox{5in}{\footnotesize\smalllineskip Table~\thetable. #1}
            \end{center}}
        \else
             {\begin{center}
             {\footnotesize Table~\thetable. #1}
              \end{center}}
        \fi}

\def\@citex[#1]#2{\if@filesw\immediate\write\@auxout
        {\string\citation{#2}}\fi
\def\@citea{}\@cite{\@for\@citeb:=#2\do
        {\@citea\def\@citea{,}\@ifundefined
        {b@\@citeb}{{\bf ?}\@warning
        {Citation `\@citeb' on page \thepage \space undefined}}
        {\csname b@\@citeb\endcsname}}}{#1}}

\newif\if@cghi
\def\cite{\@cghitrue\@ifnextchar [{\@tempswatrue
        \@citex}{\@tempswafalse\@citex[]}}
\def\citelow{\@cghifalse\@ifnextchar [{\@tempswatrue
        \@citex}{\@tempswafalse\@citex[]}}
\def\@cite#1#2{{$\null^{#1}$\if@tempswa\typeout
        {IJCGA warning: optional citation argument
        ignored: `#2'} \fi}}

\def\pmb#1{\setbox0=\hbox{#1}
        \kern-.025em\copy0\kern-\wd0
        \kern.05em\copy0\kern-\wd0
        \kern-.025em\raise.0433em\box0}

\def\fnt#1#2{\footnotetext{\kern-.3em
        {$^{\mbox{\scriptsize #1}}$}{#2}}}

\def\fpage#1{\begingroup
\voffset=.3in
\thispagestyle{empty}\begin{table}[b]\centerline{\footnotesize #1}
        \end{table}\endgroup}

\def\runninghead#1#2{\pagestyle{myheadings}
\markboth{{\protect\footnotesize\it{\quad #1}}\hfill}
{\hfill{\protect\footnotesize\it{#2\quad}}}}
\font\tenrm=cmr10
\font\tenit=cmti10
\font\tenbf=cmbx10
\font\bfit=cmbxti10 at 10pt
\font\ninerm=cmr9

\font\eightrm=cmr8

\def\qed{\hbox{${\vcenter{\vbox{                        
   \hrule height 0.4pt\hbox{\vrule width 0.4pt height 6pt
   \kern5pt\vrule width 0.4pt}\hrule height 0.4pt}}}$}}

\def\bsc{{\sc a\kern-6.4pt\sc a\kern-6.4pt\sc a}}       
\def\bflatex{\bf L\kern-.30em\raise.3ex\hbox{\bsc}\kern-.14em
T\kern-.1667em\lower.7ex\hbox{E}\kern-.125em X}

\begin{document}

\runninghead{The completeness in cumulant expansion} {The completeness in
cumulant expansion}

\normalsize\textlineskip

\thispagestyle{empty}
\setcounter{page}{1}

\copyrightheading{}                

\vspace*{0.88truein}

\fpage{1}
\centerline{\bf THE CUMULANT EXPANSION FOR THE ANDERSON LATTICE}
\vspace*{0.035truein}
\centerline{\bf WITH FINITE U: THE COMPLETENESS PROBLEM }
\vspace*{0.37truein}
\centerline{\footnotesize M. E. FOGLIO\footnote
{Associate Member of ICTP,Trieste,Italy}}
\vspace*{0.015truein}
\centerline{\footnotesize\it Instituto de F\'{\i}sica ``Gleb Wataghin'',
Universidade Estadual de Campinas,UNICAMP}
\baselineskip=10pt
\centerline{\footnotesize\it 13083-970 Campinas, S\~{a}o Paulo, Brasil
\footnote{Instituto de F\'{\i}sica ``Gleb Wataghin'',
Universidade Estadual de Campinas,UNICAMP, C.P. 6165,13083-970
Campinas, S\~{a}o Paulo, Brazil}}
\vspace*{10pt}
\centerline{\normalsize and}
\vspace*{10pt}
\centerline{\footnotesize M. S. FIGUEIRA}
\vspace*{0.015truein}
\centerline{\footnotesize\it Instituto de F\'{\i}sica,
Universidade Federal Fluminense,UFF}
\baselineskip=10pt
\centerline{\footnotesize\it 24210-340 Niter\'{o}i, Rio de Janeiro, Brasil}
\vspace*{0.225truein}
\publisher{(received date)}{(revised date)}

\vspace*{0.21truein}
\abstracts{``Completeness'' (i.e. probability conservation)
is not usually satisfied in the cumulant expansion of the Anderson
lattice when a reduced state space is employed for
$U\rightarrow \infty $. To understand this result, the well known
``Chain'' approximation is first calculated for finite $U$, followed
by taking $U\rightarrow \infty $. Completeness is recovered
by this procedure, but this result hides a serious inconsistency that
causes completeness failure in the reduced space calculation.
Completeness is satisfied and the inconsistency is removed by choosing
an adequate family of diagrams. The main result of this work is that
using a reduced space of relevant states is as good as using the whole
space.}{}{}




\vspace*{1pt}\textlineskip      
\section{Introduction}          
\vspace*{-0.5pt}
\noindent
With the Hubbard operators\cite{Hubbard5} one can describe complex
local states, as well as making substantial simplifications in the
study of configurations with a very large number of states by
considering only those few states that are relevant to the
problem\cite{Foglio}. The cumulant expansion with Hubbard operators
has been employed to calculate approximate Green's functions (GF),
both for the Hubbard model\cite{Hubbard5,CracoG} and for the
Anderson lattice\cite{Hewson,FFM}, but probability is not conserved
in the Anderson lattice when the state with two local electrons at
the same site is projected out of the four dimensional space of the
local states at each site in the infinite repulsion case. As this
problem does not appear in the Hubbard model when the whole space
is considered\cite{CracoG}, it is important to find out whether
this is an essential consequence of reducing the space. By
considering one of the simplest family of diagrams in the cumulant
expansion, we can explicitly show that the basic problem is caused
by an inconsistency that is already present in the full space
calculation. The main conclusion of the present work is that the
results obtained by employing a given family of diagrams in the
cumulant expansion would not be worse because of the space
reduction technique employed in conjunction with the Hubbard
operators. Although the space reduction is not essential in the
Anderson lattice, our result gives us confidence in the use of that
technique when the study of the full space is not viable (e.g. with
ions like $Eu^{3+}$ that has a ground $f^{7}$ configuration with
3432 states).

The Anderson lattice is an important tool in the study of strongly
correlated systems, and there are several reviews devoted to this and
closely related problems \cite{Reviews}. The model consist of a lattice with
two localized electronic states at each site, that are correlated by a
strong Coulomb repulsion $U$, plus a band of uncorrelated conduction
electrons (c-electrons) that hybridize with the localized electrons
(f-electrons).

The cumulant expansion has been employed by several authors to study the
Ising and the Heisenberg models \cite{Wortis,Englert,StinchcombeHEB}, while
Hubbard \cite{Hubbard5} extended the method to a quantum problem with
fermions. He applied the cumulant approach to the model of correlated
electrons which he had created, and studied with a different technique \cite
{Hubbard123}. The method he employed consisted in a perturbative expansion
around the atomic limit, using the hopping as perturbation, and introducing
the Hubbard operators $X_{j,ab}=\mid j,b\rangle \langle j,a\mid $ which
transform the local state $\mid a\rangle $ at site $j$ into the local state $%
\mid b\rangle $ at the same site. These X-operators make it possible to
describe in a simple way the projection of a system into the
subspace of the more relevant states, but they do not satisfy the
usual commutation properties of Bose or Fermi operators.
Nevertheless, it is possible with the use of cumulants, to derive a
diagrammatic expansion involving unrestricted lattice sums of
connected diagrams, which satisfies a linked cluster theorem
\cite{Hubbard5}. Other expansions with cumulants have been also
applied to study fundamental problems in solid state physics \cite
{Fulde,IzyumovLSBC,BartkowiakC}, and the relation of some of those
treatments to the one presented here will be briefly commented in
the final conclusions. The technique employed by Hubbard\cite
{Hubbard5,Metzner} introduces Grassman external fields without
physical meaning (unlike the magnetic fields of the corresponding
Ising expansions), but this method seems to be the natural
extension of the usual expansion with Fermi operators\cite{FFM}.

In the present work we shall consider a cumulant expansion of the
Anderson lattice, which is an extension\cite{Hewson,FFM} of the
technique employed by Hubbard to study his model. The hopping of
the conduction electrons as well as the intra-site Coulomb
repulsion $U$ between the localized (or f-electrons) are included
in the unperturbed Hamiltonian, and the hybridization is employed
as a perturbation. At a given site j, the state space of the
f-electrons is spanned by four states: the vacuum state $\mid
j,0\rangle ,$ the two states$\mid j,\sigma \rangle $ of one
f-electron with spin component $\sigma \hbar /2$ and the state
$\mid j,d\rangle $ with two electrons of opposite spin. In the
limit of infinite electronic repulsion ($U\rightarrow
\infty $) the state $\mid j,d\rangle $ is empty, and one can then consider a
reduced space of states by projecting $\mid j,d\rangle $ out. In this space,
the identity $I_{j}$ at site j should then satisfy the completeness relation:

\begin{equation}
X_{j,00}+X_{j,\sigma \sigma }+X_{j,\overline{\sigma }\overline{\sigma }%
}=I_{j}\qquad ,  \label{E1.1}
\end{equation}
where $\overline{\sigma }=-$ $\sigma $, and the three $X_{j,aa}$ are
projectors into the three states $\mid j,0\rangle $ and $\mid j,\sigma
\rangle $ of the basis. Because of the translational invariance, the
statistical averages of $X_{j,aa}$ (denoted by $n_{a})$ are independent of
j, and from Eq.~(\ref{E1.1}) they should satisfy
\begin{equation}
n_{0}+n_{\sigma }+n_{\overline{\sigma }}=1\qquad ,  \label{E1.2}
\end{equation}
a relation called ``completeness'' in what follows. It has been
found that this relation is not usually satisfied when $\mid
j,d\rangle $ is projected out of the space of local electrons and
the $n_{a}$ are calculated with approximate Green's functions
(GF)\cite{FFM}. A simple approximation displaying this behavior is
the ``Chain Approximation'' (CHA) \cite {Hewson,Enrique,FFM2},
which is the more general cumulant expansion with only second order
cumulants. This approximation is $\Phi -$derivable\cite {FFM2}, and
has other interesting properties further commented in the Appendix.
To better understand the problem, we have studied the system for
finite $U$ and then taking $U\rightarrow \infty $, a procedure that
will be called CHU in the present work. As in the case of the
Hubbard model in the full space\cite{CracoG}, the probability is
conserved in the Anderson lattice by this procedure. The analysis
of the method shows that to secure the completeness, the CHU has to
employ different values for the probabilities of the two states
$\mid j,\sigma \rangle $ and $\mid j,%
\overline{\sigma }\rangle $, which should be equal in the paramagnetic
state, and this inconsistency is what causes the lack of completeness found
with the CHA in the reduced space. This is the basic result of the present
work, that shows that the calculation in the full space satisfies
completeness through an important inconsistency, that is hidden in the
formalism.

A systematic way of adding a set of diagrams to an arbitrary family, so that
completeness be satisfied by the $n_{a}$ calculated with the corresponding
GF, has been proposed and verified in a number of cases \cite{FFM2}. When
applied to the CHA, this conjecture gives the ``Complete Chain
Approximation'' (CCHA) which satisfies Eq. (\ref{E1.2}) but it is not $\Phi
- $derivable any more. It seems interesting to repeat the calculation for
the diagrams of the CCHA employing a finite $U$ and keeping $\mid j,d\rangle
$ in the space of local states, and finally taking $U\rightarrow \infty $:
this procedure shall be called CCHU in what follows. One finds that
completeness is also satisfied by the CCHU, and that the occupation numbers
of \hspace{0pt}the two states $\mid j,\sigma \rangle $ and $\mid j,\overline{%
\sigma }\rangle $ are equal, and coincide with the value calculated directly
for the CCHA. This last result, as well as the details of the calculation,
show that the procedure of making simplified models by projecting unoccupied
states out of the space of local states, can give adequate results.

\section{The Anderson Lattice with Finite $U$}
\vspace*{-0.5pt}
\noindent
The Anderson lattice with finite $U$ is described by the Hamiltonian
\begin{eqnarray}
H &=&\sum_{\bf{k},\sigma }\ E_{\bf{k},\sigma }\
C_{\bf{k},\sigma }^{\dagger }C_{\bf{k},\sigma }+\sum_{j\sigma
}E_{j,\sigma }\ f_{j\sigma }^{\dagger }f_{j\sigma }+\sum_{j}U\
f_{j\sigma }^{\dagger }f_{j\sigma }f_{j%
\overline{\sigma }}^{\dagger }f_{j\overline{\sigma }}+  \nonumber \\
&&\sum_{jk\sigma }\left( V(k)\exp \left( i\bf{k.{R}_{j}}\right) \
f_{j\sigma }^{\dagger }C_{k\sigma }\ +H.C.\right) \ ,\   \label{E2.1}
\end{eqnarray}
where $C_{k\sigma }^{\dagger }$ ($C_{k\sigma }$) is the usual
creation (destruction) operator of conduction band electrons with
wavevector $\bf{ k}$ and spin component $\sigma \hbar /2$, and the
$f_{j\sigma }^{\dagger }$ ( $f_{j\sigma }$) are those
corresponding to the local (f) electrons at site $ j $. To make
the connection with the Hubbard operators one should substitute
\begin{equation}
f_{j\sigma }=X_{j,0\sigma }+\sigma X_{j,\overline{\sigma }d}  \label{E2.1a}
\end{equation}
into Eq. (\ref{E2.1}), where the factor $\sigma =\pm 1$ is necessary to
obtain the correct phase of the states. As the present treatment employs the
Grand Canonical Ensemble of electrons it is convenient to introduce
\begin{equation}
{\cal H}=H-\mu \left\{ \sum_{{\bf k},\sigma }C_{{\bf k},\sigma }%
^{\dagger }C_{{\bf k},\sigma }+\sum_{ja}\nu _{a}X_{j,aa}\right\}
\qquad ,
\label{E2.2}
\end{equation}
\noindent where $X_{j,aa}$ is the occupation number operator of state $\mid
a>$ at site $j$, and $\nu _{a}$ is the number of electrons in that state.
The exact and unperturbed averages of the operator $A$ are respectively
denoted by $<A>_{\cal{H}}$ and $<A>$, and it is also convenient to
introduce
\begin{equation}
\varepsilon _{j,a}=E_{j,a}-\mu \nu _{a}  \label{E2.3a}
\end{equation}
\noindent and
\begin{equation}
\varepsilon _{\bf{k\sigma }}=E_{\bf{k\sigma }}-\mu \ ,  \label{E2.3b}
\end{equation}
because the energies $E_{j,a}$ of all ionic states $\mid a>$ and
the energies $E_{\bf{k,\sigma }}$ of the conduction electrons
appear only in that form in all the calculations.

Operators $X_{j,ab}$ do not satisfy the usual anticommutation (commutation)
properties of the Fermi (Bose) operators when the two operators are at the
same site, but in this case it is sufficient to employ the product rules
\begin{equation}
X_{j,ab}\ X_{j,cd}=\delta _{b,c}\ X_{j,ad}\qquad .  \label{E2.3c}
\end{equation}
One has more freedom to define the operation between pairs of operators when
they are at different sites, and it is convenient to choose properties as
close as possible to those of the usual Fermi or Bose operators. It is then
convenient to say that $X_{j,ab}$ is of the ``Fermi type'' (``Bose type'')
when the number of electrons in the two states $\mid j,a\rangle $ and $\mid
j,b\rangle $ differ by an odd (even) number. The best choice is then to use
anticommutation relations when the two operators at different sites are of
the ``Fermi type'' and commutation relations otherwise.

The Anderson lattice has been studied in the case of $U\rightarrow \infty $
resorting to a cumulant expansion \cite{FFM} that employs the last term in $%
H $ (hybridization) as a perturbation, and this treatment is an extension of
the expansion already employed by Hubbard for his model \cite
{Hubbard5,Hubbard123}. The Hubbard operators $X_{j,ab}$ were used in this
limit to eliminate the doubly occupied states $\mid j,d\rangle $ from the
state space of the f-electrons at site $j$. The projection of $H$ into the
reduced space is

\begin{eqnarray}
H_{r} &=&\sum_{\bf{{k},\sigma }}\ E_{\bf{{k},\sigma }}\ C_{\bf{{k%
},\sigma }}^{\dagger }C_{\bf{{k},\sigma }}+\sum_{j,\sigma }\
E_{j,\sigma }\ X_{j,\sigma \sigma }  \nonumber
\\ &&+\sum_{j,\bf{{k},\sigma }}\left( V_{j,\bf{{k},\sigma
}}\ X_{j,0\sigma }^{\dagger }\ C_{\bf{{k},\sigma
}}+V_{j,\bf{{k},\sigma } }^{*}\ C_{\bf{{k},\sigma }}^{\dagger }\
X_{j,0\sigma }\right) \qquad .
\label{E2.4}
\end{eqnarray}

The method has been generalized to several configurations with a rather
arbitrary choice of states \cite{FFM}, and can be used to study the Anderson
lattice with finite $U$, described by the Hamiltonian in Eq. (\ref{E2.1}).
The cumulant expansion gives the GF of the Matsubara type for imaginary time
$\tau $%
\begin{equation}
\left\langle \left( \widehat{X}_{j,\alpha }(\tau )\ \widehat{X}_{j^{\prime
},\alpha ^{\prime }}\right) _{+}\right\rangle _{\cal{H}}\qquad ,
\label{E2.4a}
\end{equation}
where $\widehat{X}_{j,\alpha }(\tau )=\exp \left( \tau \cal{H}\right)
X_{j,\alpha }\exp \left( -\tau \cal{H}\right) $ corresponds to the
Heisenberg representation, the subindex $\alpha =(b,a)$ represents the
transition $\mid a>\rightarrow \mid b>$ and the subindex $+$ indicates that
the operators inside the parenthesis are taken in the order of increasing $%
\tau $ to the left, with a change of sign when the two Fermi-type operators
have to be exchanged to obtain this ordering.

The GFs in Eq. (\ref{E2.4a}) are defined in the interval $0\leq \tau \leq
\beta \equiv 1/T$ and, because of their boundary condition in this variable%
\cite{FFM}, one can associate to them a Fourier series with a
coefficient for each Matsubara frequency $\omega _{\nu }=\pi \nu
/\beta $ (where $\nu $ are the even or odd integers, depending on
whether the operator $X_{\alpha }$ is of the Bose type or of the
Fermi type respectively). One can also transform the GF to
reciprocal space \cite{FFM} and, because of the invariance against
time and lattice translations, they are proportional to quantities
$G_{\alpha \alpha ^{\prime }}^{ff}(\bf{k},\omega _{\nu })$. One can
also associate these quantities to the $i\omega _{\nu }$ points of
a complex plane in the variable $z=\omega +iy$ and, in the usual
way \cite {Negele}, one can make the analytic continuation to the
upper and lower half-planes of the complex frequency $z$, obtaining
a function $\overline{G}%
_{\alpha \alpha ^{\prime }}^{ff}(\bf{k},z)$ which is minus the Fourier
transform of the real time GF. By an adequate choice of $\alpha $ and $%
\alpha ^{\prime }$ it is possible to use these functions to calculate the
occupation $n_{a}$ of the local state $\mid a\rangle :$%
\begin{equation}
n_{a}=\int_{-\infty }^{\infty }\rho _{ba}(\omega )\ f_{T}(\omega )\ d\omega
\qquad ,  \label{E2.5}
\end{equation}
\noindent where
\begin{equation}
f_{T}(z)=\left\{ 1+\exp \left( \beta \ z\right) \right\} ^{-1}  \label{2.5a}
\end{equation}
is the Fermi function and
\begin{equation}
\rho _{ba}(\omega )=\frac{1}{\pi }\ \lim_{\epsilon \rightarrow 0}\ {\it Im%
}\left\{ \frac{1}{N_{s}}\sum_{\bf{k}}\overline{G}_{ba,ab}^{ff}(\bf{k}%
,\omega +i\left| \epsilon \right| )\right\}  \label{2.6}
\end{equation}
is the spectral density associated to the transition $\alpha =(b,a)$. With
the same $\rho _{ba}(\omega )$ it is also possible to obtain the occupation $%
n_{b}$ of the local state $\mid b\rangle :$%
\begin{equation}
n_{b}=\int_{-\infty }^{\infty }\rho _{ba}(\omega )\ (1-f_{T}(\omega ))\
d\omega \qquad .  \label{E2.7}
\end{equation}
When the state $\mid d>$ is projected out, the occupation $n_{\sigma }$ can
be calculated with only Eq. (\ref{E2.5}) and $\overline{G}_{0\sigma ,\sigma
0}^{ff}$, while in the whole space one could also use Eq. (\ref{E2.7}) and $%
\overline{G}_{\sigma d,d\sigma }^{ff}$. These two values of $n_{\sigma }$
are not automatically equal when one uses approximate GFs, and this property
is the origin of the results obtained in the present work.
\vspace*{12pt}

\subsection{The chain approximation (CHA)}
\label{SC}
\noindent
The family of diagrams which give this approximation is shown in figure \ref
{F1}. In the diagrams of the cumulant expansion \cite{FFM}, the circles
(vertices) correspond to cumulants: the filled ones for the f-electrons and
the empty ones for the c-electrons. In the present expansion the
hybridization is the perturbation, and it is represented in the diagrams by
the lines (edges) joining two vertices. Because of the form of the
hybridization in Eq. (\ref{E2.1}), the edges can only join a c-vertex to an
f-vertex. The number of edges reaching a vertex gives the order of the
cumulant, and only those of second order appear in the CHA. All the
cumulants of second order coincide with the free propagators, and because of
Wick's theorem, these are the only non-zero cumulants of the c-electrons,
while there are cumulants of any even order for the f-electrons, which can
be calculated employing a generalized version of Wick's theorem \cite
{Hewson,FFM2,YangW}

\begin{figure}[H]
\caption[Fig.1]{The diagrams for the GF in the chain approximation (CHA).
The circles (vertices) correspond to cumulants: the filled ones for the
f-electrons and the empty ones for the c-electrons. The perturbation
(hybridization) is represented in the diagrams by the lines (edges) joining
two vertices. a) The CHA diagrams for the f-electrons GF, represented by the
filled square to the right. b) Same as a) for the c-electrons, represented
by an empty square. }
\label{F1}
\end{figure}

\vspace*{12pt}
\subsubsection{The Anderson lattice for $U\rightarrow \infty $ in the
reduced space}
\label{SC.1}
\noindent
The f-electron GF has been calculated for the CHA \cite{Physica} in the
reduced space, and using the simpler notation $G_{0\sigma ,\sigma 0}(\bf{%
k},z)\equiv \overline{G}_{0\sigma ,\sigma 0}^{ff}(\bf{k},z)$ we can write
\begin{equation}
G_{0\sigma ,0\sigma }(\bf{k},z)=-\frac{D_{0\sigma }^{0}(z-\varepsilon _{%
\bf{k}\sigma })}{(z-\varepsilon _{1}(\bf{k}))(z-\varepsilon _{2}(%
\bf{k}))}\qquad .  \label{E3.1}
\end{equation}
The energies $\varepsilon _{1}(\bf{k})$ and $\varepsilon _{2}(\bf{k}%
) $ are those of the two elementary excitations with wave vector $\bf{k}$%
, resulting from the hybridization of a band $\varepsilon _{\bf{k}\sigma
}$ and a dispersionless band of energy $\varepsilon _{f}=\varepsilon _{{j}%
\sigma }$ with a reduced hybridization constant $\sqrt{D_{0\sigma }^{0}}V(k)$%
. They are given by the two roots of $(z-\varepsilon _{f})(z-\varepsilon _{%
\bf{k}\sigma })-D_{0\sigma }^{0}\left| V(k)\right| ^{2}=0$, where
\begin{equation}
D_{0\sigma }^{0}=\left\langle X_{00}+X_{\sigma \sigma }\right\rangle \qquad
\label{E3.3}
\end{equation}
and $\left\langle A\right\rangle $ is the unperturbed average of $A$. This
GF is equal to that of two uncorrelated bands of energies $\varepsilon _{f}$
and $\varepsilon _{\bf{k}\sigma }$ hybridized with a parameter $\sqrt{%
D_{0\sigma }^{0}}\ V(k)$, showing that the only effect of the correlations
in the CHA is to reduce the hybridization by a factor $\sqrt{D_{0\sigma }^{0}%
}$. A typical spectral density $\rho _{0\sigma }(\omega )$ is shown in
figure \ref{F2}, and it is easy to show that
\begin{equation}
n_{0}+n_{\sigma }=\int\nolimits_{-\infty }^{\infty }d\omega \ \rho _{0\sigma
}(\omega )=D_{0\sigma }^{0}\qquad ,  \label{E3.3a}
\end{equation}
i.e. $n_{0}+n_{\sigma }$, which is the total area of the spectral density in
the CHA, coincides with the unperturbed value of the same quantity.

\begin{figure}[H]
\caption[Fig.2]{Plot of the spectral density of f-electrons for the CHA as a
function of $\omega $ for a rectangular band of total width $2W=\pi $
centered at $E=0$, with single local electron energy $E_{j,\sigma }=-0.5$
equal to the chemical potential $\mu $, and with hybridization parameter $%
V=0.3$, all measured in the same energy units. A gap appears at $\mu \approx
\varepsilon _{f}=E_{j,\sigma }-\mu $.}
\label{F2}
\end{figure}

\vspace*{12pt}

\subsubsection{The Anderson lattice for $U\rightarrow \infty $ in the full
space}
\label{SC.2}
\noindent
For the Anderson lattice with finite $U$, described by the Hamiltonian in
Eq. (\ref{E2.1}), the same family of diagrams shown in figure \ref{F1}
corresponds to the CHA. One can employ the generalized method introduced in
ref. \cite{FFM} for a rather arbitrary choice of states, and obtain the
approximate GFs for finite $U$. In reciprocal space and for the complex
frequency $z$ they are
\begin{equation}
G_{0\sigma ,0\sigma }(k,z)=-\frac{D_{0\sigma }^{0}(1-b)}{(z-\varepsilon
_{f})(1-a-b)}\quad ,  \label{E3.4}
\end{equation}
\begin{equation}
G_{\overline{\sigma }d,0\sigma }(k,z)=-\frac{D_{\overline{\sigma }d}^{0}\
\sigma \ a}{(z-\varepsilon _{f}-U)(1-a-b)}\quad ,  \label{E3.5}
\end{equation}
\begin{equation}
G_{0\sigma ,\overline{\sigma }d}(k,z)=-\frac{D_{0\sigma }^{0}\ \sigma \ b}{%
(z-\varepsilon _{f})(1-a-b)}\quad ,  \label{E3.6}
\end{equation}
\begin{equation}
G_{\overline{\sigma }d,\overline{\sigma }d}(k,z)=-\frac{D_{\overline{\sigma }%
d}^{0}\ (1-a)}{(z-\varepsilon _{f}-U)(1-a-b)}\quad ,  \label{E3.7}
\end{equation}
where
\begin{equation}
a=\frac{D_{0\sigma }^{0}\left| V(k)\right| ^{2}}{(z-\varepsilon
_{f})(z-\varepsilon _{\bf{k}\sigma })}\qquad ,  \label{E3.7a}
\end{equation}
\begin{equation}
b=\frac{D_{\overline{\sigma }d}^{0}\left| V(k)\right| ^{2}}{(z-\varepsilon
_{f}-U)(z-\varepsilon _{\bf{k}\sigma })}\qquad ,  \label{E3.7b}
\end{equation}
\begin{equation}
D_{\overline{\sigma }d}^{0}\ =\left\langle X_{\overline{\sigma }\overline{%
\sigma }}+X_{dd}\right\rangle \quad ,  \label{E3.7c}
\end{equation}
and $D_{0\sigma }^{0}$ is given in Eq. (\ref{E3.3}). However, it must be
noted that now we are employing $<A>$ to indicate the unperturbed average of
$A$ in the full space, i.e. for the Hamiltonian of Eq. (\ref{E2.1}) taken
with $V(k)=0$.

The identity in the full space is given by
\begin{equation}
X_{j,00}+X_{j,\sigma \sigma }+X_{j,\overline{\sigma }\overline{\sigma }%
}+X_{j,dd}=I_{j}\qquad  \label{E3.7cd}
\end{equation}
rather than by Eq. (\ref{E1.1}), and the completeness with finite $U$ then
becomes
\begin{equation}
n_{0}+n_{\sigma }+n_{\overline{\sigma }}+n_{d}=1\qquad .  \label{E3.7d}
\end{equation}
Note that completeness is automatically satisfied for the unperturbed case,
because then $n_{0}+n_{\sigma }+n_{\overline{\sigma }}+n_{d}=D_{0\sigma
}^{0}+D_{\overline{\sigma }d}^{0}=1$, and it is worth analyzing the two
limit cases $U=0$ and $U\rightarrow \infty $ for the chain approximation.
\vspace*{12pt}
\paragraph{The $U=0$ case}
\label{SC3.2.a}
\noindent
Employing
\begin{equation}
<<\ f_{j\sigma };\ f_{j\sigma }^{\dagger }>>=G_{0\sigma ,0\sigma }+\sigma G_{%
\overline{\sigma }d,0\sigma }+\sigma G_{0\sigma ,\overline{\sigma }d}+G_{%
\overline{\sigma }d,\overline{\sigma }d}\qquad ,  \label{E3.7e}
\end{equation}
we obtain the exact GF in this limit, and completeness is satisfied.
Although this result is not surprising, it serves to illustrate the nature
of the $X$ operators, which contain strong correlations as a result of their
own definition. This is clearly shown by employing Eqs. (\ref{E3.5},\ref
{E3.6}) to calculate $\left\langle X_{j,0\sigma }X_{j,\overline{\sigma }%
d}\right\rangle $ and $\left\langle X_{j,\overline{\sigma }d}X_{j,0\sigma
}\right\rangle $: these averages should be zero because of the rules of
product of the $X$ operators (i.e. Eq. (\ref{E2.3c})), but a different value
is obtained in the CHA. This result can be interpreted as follows: Wick's
theorem is satisfied by the Fermion operators $f_{j\sigma }$ and $\
f_{j\sigma }^{\dagger }$, and all the cumulants of order higher than two are
zero, so that the CHA gives the exact solution for these operators. As
Wick's theorem is not satisfied by the $X$ operators, all the diagrams with
cumulants higher than two are missing from the individual GF in Eqs. (\ref
{E3.4} -\ref{E3.7}), and they are not exact in the CHA. It is by a
cancellation automatically built in the formalism that Eq. (\ref{E3.7e})
gives the exact $<<f_{j\sigma };\ f_{j\sigma }^{\dagger }>>$, although the $%
G_{\alpha \alpha ^{\prime }}^{ff}$ are only approximate.

The relation $X_{j,0\sigma }X_{j,\overline{\sigma }d}=0$ is an
example of the ``kinematic restriction which forbids more than one
energy level of a single ion be occupied'' \cite{YangW75}. The
failure of that restriction to be satisfied in the average by many
approximate GF of the Heisenberg and related models is discussed in
\cite{YangW75}. A cure of this problem for the Heisenberg model
with uniaxial anisotropy is proposed in that reference, and we
shall later return to this property in connection with our problem.
\vspace*{12pt}
\paragraph{The $U\rightarrow \infty $ case}
\label{SC3.2.b}
\noindent
To study this limit one should start with a very large $U$. In the usual
case, the band of conduction electrons is close to $\varepsilon _{f}$, and
from Eq. (\ref{E3.7b}) it is $b\rightarrow 0$ when $U\rightarrow \infty .$ A
fairly symmetrical situation is obtained when the conduction band is close
to $\varepsilon _{f}+U$, and in this case it follows that $a\rightarrow 0$
when $U\rightarrow \infty $. When both $\varepsilon _{f}$ and $\varepsilon
_{f}+U$ are far from the conduction band and $U\rightarrow \infty $, both $%
b\rightarrow 0$ and $a\rightarrow 0,$ and the GF coincide with those of the
unperturbed problem. In all cases the two ``cross'' GFs, i.e. $G_{\overline{%
\sigma }d,0\sigma }$ and $G_{0\sigma ,\overline{\sigma }d}$ of Eqs. (\ref
{E3.5}, \ref{E3.6}), vanish for any value of $z$ when $U\rightarrow \infty ,$
so that the kinematic restriction discussed in reference \cite{YangW75} is
automatically satisfied in this limit of the CHA.

Let us consider only the case when $\varepsilon _{f}$ is close to the
conduction electron band. The Eq. (\ref{E3.4}) then becomes
\begin{equation}
G_{0\sigma ,0\sigma }(\bf{k},z)=-\frac{D_{0\sigma }^{0}(z-\varepsilon _{%
\bf{k}\sigma })}{(z-\varepsilon _{1}(\bf{k}))(z-\varepsilon _{2}(%
\bf{k}))}\qquad ,  \label{E3.8}
\end{equation}
which is just the same Eq. (\ref{E3.1}) obtained in the CHA when the state $%
\mid j,d\rangle $ with two f electrons is projected out. In the two ways of
dealing with $\mid j,d\rangle $, Eq. (\ref{E3.3a}) shows that $%
n_{0}+n_{\sigma }$ coincides with the unperturbed value $D_{0\sigma }$, and
the effect of the perturbation consist in transferring weights between $%
n_{0} $ and $n_{\sigma }$. Equation (\ref{E3.7}) becomes
\begin{equation}
G_{\overline{\sigma }d,\overline{\sigma }d}(\bf{k},z)=-\frac{D_{%
\overline{\sigma }d}^{0}}{(z-\varepsilon _{f}-U)}\qquad ,  \label{E3.9}
\end{equation}
and the corresponding $\rho _{\overline{\sigma }d}$ is a delta function with
weight $D_{\overline{\sigma }d}^{0}$ at $\varepsilon _{f}+U$. Applying Eq. (%
\ref{E2.5}) and Eq. (\ref{E2.7}) to $\rho _{\overline{\sigma }d}$ we obtain $%
n_{d}=D_{\overline{\sigma }d}^{0}\ f_{T}(\varepsilon _{f}+U)$ and $n_{%
\overline{\sigma }}=D_{\overline{\sigma }d}^{0}\ (1-f_{T}(\varepsilon
_{f}+U)).$ The relation $n_{\overline{\sigma }}+n_{d}=D_{\overline{\sigma }%
d}^{0}$ is then valid independently of the value of $\mu ,$ and together
with Eq. (\ref{E3.3a}) shows that the completeness in the full space is
satisfied: $n_{0}+n_{\sigma }+n_{\overline{\sigma }}+n_{d}=D_{0\sigma
}^{0}+D_{\overline{\sigma }d}^{0}=1$. Notice that when $\varepsilon _{f}$ is
close to the conduction electron band, $\varepsilon _{f}+U$ is very large
and $n_{d}=D_{\overline{\sigma }d}^{0}\ f_{T}(\varepsilon _{f}+U)\rightarrow
0,$ so that completeness in the reduced space $n_{0}+n_{\sigma }+n_{%
\overline{\sigma }}=1$ is also satisfied. Also the unperturbed $n_{d}^{0}=0$%
, so that from $n_{\overline{\sigma }}=D_{\overline{\sigma }d}^{0}$ follows
that $n_{\overline{\sigma }}=n_{\overline{\sigma }}^{0}$, and this value is
usually different from the $n_{\sigma }$ calculated by the CHA. This shows
that completeness is satisfied in the present calculation at the cost of
using different values for $n_{\sigma }$ and $n_{\overline{\sigma }}$. As
Eqs. (\ref{E3.8},\ref{E3.9}) are valid for the two values of $\sigma $,
different values of $n_{\sigma }$ are obtained when they are calculated with
$G_{0\sigma ,0\sigma }$ vs. $G_{\sigma d,\sigma d}$. When the state $\mid
j,d\rangle $ is projected out, the Eq. (\ref{E3.9}) cannot be used, and one
is then forced to calculate $n_{\overline{\sigma }}$ with $G_{0\overline{%
\sigma },0\overline{\sigma }}$, so that $n_{\overline{\sigma }}=n_{\sigma }$%
, and completeness can not be satisfied by the CHA in the reduced space. The
different values of $n_{\overline{\sigma }}$ and $n_{\sigma }$ obtained with
the CHU are shown in figures \ref{F3} and \ref{F4} as a function of $T$ for
the following parameters: $E_{j,\sigma }=-0.5$, a local hybridization with $%
V=0.3$ and a rectangular conduction band centered at $E_{\bf{k},\sigma
}=0$ and with a total width $2W=\pi $. The chemical potential in the two
figures is respectively $\mu =-0.60$ and $\mu =-0.40$, and the corresponding
dotted curves show the crossover, that occurs at low $T$ in the system with $%
V=0$, from the maximal occupation $n_{\sigma }^{0}=0.5$ of the local
electron to the empty state $n_{\sigma }^{0}=0$.

\begin{figure}[H]
\caption[Fig.3]{The occupation number $n_{\sigma }$ is plotted as a function
of $T$ for several different approximations and for the same parameters used
in figure~\ref{F2}, but with chemical potential $\mu =-0.60$. The dashed
line is $n_{\sigma}$ calculated with the CHU, and coincides with the same
parameter calculated with the CHA. The dotted line is $n_{\overline{\sigma }
} $ calculated with the CHU, which coincides with the unperturbed value $%
n_{\sigma }^0$ (i.e. for $V=0$). The full line is $n_{\sigma }$ calculated
with the CCHU and coincides with $n_{\overline{\sigma }}$ calculated within
the same approximation and also with the $n_{\sigma }$ obtained with the
CCHA. Note that $n_{\overline{\sigma }}$ and $n_{\sigma } $ are different
for the CHU (dotted vs. dashed line). }
\label{F3}
\end{figure}

\begin{figure}[H]
\caption[Fig.4]{Same as in figure~\ref{F3} but for $\mu =-0.40$. }
\label{F4}
\end{figure}

In many works\cite{CracoG,Metzner,BartkowiakC2} it is the GF $<<\ f_{j\sigma
};\ f_{j\sigma }^{\dagger }>>$ rather than the partial GF that is
calculated, and completeness is satisfied in the full space. As it was shown
above, this could hide the fact that there is a pair of conflicting values
of $n_{\sigma }$ for each $\sigma $, and that different values of $n_{%
\overline{\sigma }}$ and $n_{\sigma }$ are necessary in the paramagnetic
case to satisfy completeness: the relation Eq.(\ref{E3.7d} ) would
not be satisfied if only one of the two possible values were used
for both $n_{%
\overline{\sigma }}$ and $n_{\sigma }$. The correct completeness obtained
from $<<\ f_{j\sigma };\ f_{j\sigma }^{\dagger }>>$ masks this basic failure
of the CHA, and the properties calculated in the CHU would not be more
reliable than those obtained in the CHA when $\mid j,d\rangle $ is projected
out.
\vspace*{1pt}

\section{The Complete Chain Approximation (CCHA)}
\vspace*{-0.5pt}
\label{SD}
\noindent
A conjecture that gives a systematic way of achieving completeness by adding
a set of diagrams to an arbitrary family was stated in a previous paper \cite
{FFM2}, and it was verified in a number of cases that include diagrams with
infinite fourth-order cumulants, but a general derivation has not been
found. This conjecture gives for the CHA the ``Complete Chain
Approximation'' (CCHA) which satisfies Eq. (\ref{E1.2}), and the extra
diagrams are shown in figure \ref{F5}. We shall now apply to this family of
diagrams the procedure (named CCHU in this case) employed in Section~\ref
{SC.2}, of keeping $\mid j,d\rangle $ in the space of local states for
finite $U$ and then taking $U\rightarrow \infty $. As completeness in the
full space is satisfied by the CHU, one wonders whether adding the extra
diagrams to this approximation would maintain Eq. (\ref{E3.7d}). This
equation is satisfied in the CHU at the cost of having different occupation
numbers for \hspace{0pt}the two states $\mid j,\sigma \rangle $ and $\mid j,%
\overline{\sigma }\rangle $, and one also wonders whether this inconsistency
would be removed in the CCHU. In this section it is shown that the two
questions have an affirmative answer, and that the corresponding $n_{\sigma
} $ and $n_{\overline{\sigma }}$ coincide with the value calculated directly
in the CCHA.

\begin{figure}[H]
\caption[Fig.5]{a) The Complete Chain Approximation: empty squares symbolize
the diagrams of the c-electron GF in the CHA (cf. figure~\ref{F1}b). b) The
diagrams added to the CHA to make it complete. }
\label{F5}
\end{figure}

The fact that different average values of the same operator can be
obtained from different GF that correspond to the same
approximation has been well known in magnetic models
\cite{YangW75}, and in that reference they propose that when a
given correlation function is calculated, the diagrams should be
classified according to the number of momentum variables summed
over in the diagram and be consistently included in the calculation
according to this classification (see also references
\cite{Stinchcombe,CottamS}). This procedure is closely related to
the conjecture stated in \cite{FFM2}, but this last gives clear
rules for the diagrams that should be included in a GF for the
Anderson lattice as well as for other systems of fermions. The
procedure stated in \cite{YangW75} associates a given family of
diagrams to the calculation of a given average rather than to a GF:
when their prescription is applied to our problem, we have a GF in
the CHA without momentum integrations, but it is necessary to
integrate once over the wave vectors when the number of electrons
is calculated, so that further diagrams should be included. Our
conjecture seems more clear because it applies directly to the GF,
and has the same effect as the proposal in \cite{YangW75}.
\vspace*{12pt}

\subsection{The GF in the CCHU}
\label{SD.1}
\noindent
The GFs obtained for finite $U$ in the CCHA are rather complicated, and only
the corrections $\Delta G_{\alpha ,\alpha ^{\prime }}(\bf{k},z)$ that
could give a non-zero contribution to the occupation numbers when $%
U\rightarrow \infty $ are given below. The calculation of these quantities
follows the same technique employed in the Appendix C of reference \cite
{FFM2} to derive the function $S_{\sigma }^{CH}(\bf{k},\omega _{\nu })$,
which corresponds to the present $\Delta G_{0\sigma ,0\sigma }(\bf{k}%
,\omega _{\nu })$ when $U\rightarrow \infty $. We obtain:
\begin{eqnarray}
\Delta G_{0\sigma ,0\sigma }(\bf{k},\omega _{\nu }) &=&-\left|
V(k)\right| ^{2}\left\{ A\ K_{0\sigma }(\omega _{\nu })+B\ K_{0\sigma
}^{2}(\omega _{\nu })\ C_{\sigma }(\omega _{\nu })+\right.  \nonumber \\
&&\left. B^{\prime }\ K_{0\sigma }^{2}(\omega _{\nu })\ H_{0d}(\omega _{\nu
})+CK_{0\sigma }^{2}(\omega _{\nu })\right\} \ ,  \label{E4.1}
\end{eqnarray}
\begin{eqnarray}
\Delta G_{\overline{\sigma }d,\overline{\sigma }d}(\bf{k},\omega _{\nu
}) &=&-\left| V(k)\right| ^{2}\left\{ -A\ K_{\overline{\sigma }d}(\omega
_{\nu })+B\ K_{\overline{\sigma }d}^{2}(\omega _{\nu })\ C_{\sigma }(\omega
_{\nu })+\right.  \nonumber \\
&&\left. B^{\prime }\ K_{\overline{\sigma }d}^{2}(\omega _{\nu })\
H_{0d}(\omega _{\nu })-CK_{\overline{\sigma }d}^{2}(\omega _{\nu })\right\}
\ ,  \label{E4.2}
\end{eqnarray}
and we do not give the expressions for $G_{\overline{\sigma }d,0\sigma
}(k,z) $ and $G_{0\sigma ,\overline{\sigma }d}(k,z)$ because, as in the CHA,
they do not contribute to $n_{a}$ when $U\rightarrow \infty $. The following
definitions have been used:
\begin{equation}
K_{ab}(\omega _{\nu })=-\frac{1}{i\omega _{\nu }+(\varepsilon
_{a}-\varepsilon _{b})}\qquad ,  \label{E4.3}
\end{equation}
\begin{equation}
C_{\sigma }(\omega _{\nu })=\frac{1}{N}\sum_{\bf{k}}\frac{(i\omega _{\nu
}-\varepsilon _{f})}{(i\omega _{\nu }-\varepsilon _{1})\ (i\omega _{\nu
}-\varepsilon _{2})}\qquad ,  \label{E4.4}
\end{equation}
\begin{equation}
\beta H_{0d}(\omega _{\nu })=\sum_{\omega _{1}}C_{\sigma }(\omega _{1})K_{%
\overline{0}d}(\omega _{\nu }+\omega _{1})\qquad ,  \label{E4.5}
\end{equation}
and both $\varepsilon _{1}=\varepsilon _{1}(\bf{k})$ and $\varepsilon
_{2}=\varepsilon _{2}(\bf{k})$ have been given just after Eq. (\ref{E3.1}%
). We have left out from the constants in Eqs. (\ref{E4.1},\ref{E4.2}) all
those terms that would vanish when $U\rightarrow \infty $ (note in
particular that $\left\langle X_{dd}\right\rangle =0$): they can then be
expressed by $A=\left\langle X_{00}\right\rangle \ \left\langle X_{\sigma
\sigma }\right\rangle \ \beta \ I_{1}-D_{0\sigma }^{0}\ I_{2}$; $%
B=-\left\langle X_{\sigma \sigma }\right\rangle \ (1+D_{0\sigma }^{0})$; $%
B^{\prime }\ =D_{0d}^{0}=-\left\langle X_{00}\right\rangle $ and $%
C=-D_{0\sigma }^{0}\ I_{1}$, where
\begin{equation}
I_{\ell }=\frac{1}{\beta }\sum_{\omega _{1}}C_{\sigma }(\omega _{1})\
K_{\sigma }^{\ell }(\omega _{1})\qquad \rm{(with\quad \ell =1,2)\qquad }.
\label{E4.6}
\end{equation}
These values of $A$, $B$ and $C$ are equal\footnote{%
But note that there is a typing mistake in that reference: the expression
for $B$ should read $B=-(D_{\sigma }^{0}\ (1-D_{\sigma }^{0})+{\it x}%
_{\sigma })$.} to the parameters with the same name in Eq. (4.5) of
reference \cite{FFM2}, while the parameter $B^{\prime }$ only appears in the
present treatment. The terms with $A$, $B$ and $C$ in Eqs. (\ref{E4.1},\ref
{E4.2}) are of the same type of the corresponding ones in reference \cite
{FFM2}, and the usual analytic continuation from their values at all the
points $\varsigma _{\nu }=i\omega _{\nu }$to the complex plane $\varsigma $
give analytic functions off the real axis.

The sum over $\omega _{1}$ in Eq. (\ref{E4.5}) was obtained by the usual
technique, and the resulting $H_{0d}(\omega _{\nu })$ has several terms in
which $\ f_{T}(\varepsilon _{d}-i\ \omega _{\nu })$ is present. To obtain an
analytic function off the real axis by analytic continuation of the values
of $H_{0d}(\omega _{\nu })$, it is necessary to use the following property
of $f_{T}(z)$:
\begin{equation}
f_{T}\left( \varepsilon _{d}-i\ \omega _{\nu }\right) =\left( 1-\exp \left(
\beta \varepsilon _{d}\right) \right) ^{-1}\equiv -b_{T}\left( \varepsilon
_{d}\right) \qquad ,  \label{E4.7}
\end{equation}
where $b_{T}\left( z\right) $ is the Bose equivalent to the $f_{T}\left(
z\right) $. It then follows that
\begin{eqnarray}
H_{0d}(\omega _{\nu }) &=&\frac{f_{T}\left( \varepsilon _{1}\right)
-f_{T}\left( \varepsilon _{2}\right) }{\varepsilon _{1}-\varepsilon _{2}}
\nonumber \\
&&-\frac{i\ \omega _{\nu }-\varepsilon _{d}+\varepsilon _{f}}{\varepsilon
_{1}-\varepsilon _{2}}\left\{ \frac{f_{T}\left( \varepsilon _{1}\right)
+b_{T}\left( \varepsilon _{d}\right) }{i\ \omega _{\nu }-\varepsilon
_{d}+\varepsilon _{1}}-\frac{f_{T}\left( \varepsilon _{2}\right)
+b_{T}\left( \varepsilon _{d}\right) }{i\ \omega _{\nu }-\varepsilon
_{d}+\varepsilon _{2}}\right\}  \nonumber \\
&&+b_{T}\left( \varepsilon _{d}\right) \frac{i\ \omega _{\nu }-\varepsilon
_{d}+\varepsilon _{f}}{\left( i\ \omega _{\nu }-\varepsilon _{d}+\varepsilon
_{1}\right) \left( i\ \omega _{\nu }-\varepsilon _{d}+\varepsilon
_{2}\right) }\qquad ,  \label{E4.8}
\end{eqnarray}
and the analytic continuation of the total GF would also be an analytic
function off the real axis, making it possible to employ Eqs. (\ref{E2.5}-%
\ref{E2.7}) to calculate the occupation numbers.
\vspace*{12pt}

\subsection{The occupation numbers for the Anderson lattice in the CCHU}
\label{SD.2}
\noindent
It is possible to obtain partial results that are fairly independent of the
system parameters by integrating Eqs. (\ref{E2.5},\ref{E2.7}) over $\omega $
for each value of $\bf{k}$, and leaving the sum over this variable as a
final step. One calculates the occupation number $n_{a}(\bf{k})$
associated to the wavevector $\bf{k}$ and the state $\mid a\rangle $ by
substituting the local spectral density $\rho _{ba}(\omega )$ in Eq. (\ref
{E2.5}) or Eq. (\ref{E2.7}) by
\begin{equation}
\rho _{ba}(\bf{k,}\omega )=\frac{1}{\pi }\ \lim_{\epsilon \rightarrow
0}\ {\it Im}\left\{ \overline{G}_{ba,ab}^{ff}(\bf{k},\omega +i\left|
\epsilon \right| )\right\} \qquad .  \label{E4.9}
\end{equation}
Defining $D_{0\sigma }(\bf{k})=n_{0}(\bf{k})+n_{\sigma }(\bf{k})$
and $D_{\overline{\sigma }d}(\bf{k})=n_{\overline{\sigma }}(\bf{k}%
)+n_{d}(\bf{k})$ we obtain that
\begin{equation}
D_{0\sigma }(\bf{k})=D_{0\sigma }^{0}-\left| V(k)\right| ^{2}\ A
\label{E4.10}
\end{equation}
and
\begin{equation}
D_{\overline{\sigma }d}(\bf{k})=D_{\overline{\sigma }d}^{0}+\left|
V(k)\right| ^{2}\ A\qquad ,  \label{E4.11}
\end{equation}
so that
\begin{equation}
n_{0}(\bf{k})+n_{\sigma }(\bf{k})+n_{\overline{\sigma }}(\bf{k}%
)+n_{d}(\bf{k})=D_{0\sigma }^{0}+D_{\overline{\sigma }d}^{0}=1\qquad .
\label{E4.12}
\end{equation}
The sum of Eq. (\ref{E4.12}) over all the $N$ values of $\bf{k}$ divided
by $N$ shows that completeness (i.e. Eq.(\ref{E3.7d})) is satisfied by any
set of system parameters in the CCHU, in particular by any unperturbed
density $\rho _{c}^{0}(\varepsilon )$ of c-electron states with respect to
their energy and by any $V(k)$.\newline
Using Eqs. (\ref{E3.8},\ref{E4.1}) we find that
\begin{eqnarray}
n_{\sigma }(\bf{k}) &=&D_{0\sigma }^{0}\left\{ \frac{\varepsilon
_{1}f_{T}\left( \varepsilon _{1}\right) -\varepsilon _{2}f_{T}\left(
\varepsilon _{2}\right) }{\varepsilon _{1}-\varepsilon _{2}}-\varepsilon _{%
\bf{k\sigma }}\ \frac{f_{T}\left( \varepsilon _{1}\right) -f_{T}\left(
\varepsilon _{2}\right) }{\varepsilon _{1}-\varepsilon _{2}}\right\}
\label{E4.13} \\
&&-\left| V(k)\right| ^{2}\left\{ A\ f_{T}(\varepsilon _{f})+B\
g(\varepsilon _{\bf{k\sigma }})+C\beta f_{T}(\varepsilon _{f})\left(
1-f_{T}(\varepsilon _{f})\right) \right\} \qquad ,  \nonumber
\end{eqnarray}
where
\begin{equation}
g(\varepsilon _{\bf{k\sigma }})=\frac{f_{T}(\varepsilon _{f})}{%
(\varepsilon _{f}-\varepsilon _{1})\ (\varepsilon _{f}-\varepsilon _{2})}+%
\frac{f_{T}\left( \varepsilon _{1}\right) }{(\varepsilon _{1}-\varepsilon
_{f})\ (\varepsilon _{1}-\varepsilon _{2})}+\frac{f_{T}\left( \varepsilon
_{2}\right) }{(\varepsilon _{2}-\varepsilon _{f})\ (\varepsilon
_{2}-\varepsilon _{1})}\qquad ,  \label{E4.14}
\end{equation}
and employing Eq. (\ref{E4.10}) one obtains
\begin{equation}
n_{0}(\bf{k})=D_{0\sigma }-n_{\sigma }(\bf{k})\qquad .
\label{E4.14a}
\end{equation}
From Eqs. (\ref{E3.9},\ref{E4.2}) follows that $n_{d}(\bf{k})=0$, and
therefore
\begin{equation}
n_{\overline{\sigma }}(\bf{k})=D_{\overline{\sigma }d}(\bf{k})=D_{%
\overline{\sigma }d}^{0}+\left| V(k)\right| ^{2}\ A\qquad .  \label{E4.15}
\end{equation}

From these expressions is not obvious a priori that one would obtain $n_{%
\overline{\sigma }}=n_{\sigma }$ after summing the Eqs. (\ref{E4.13},\ref
{E4.15}) over $\bf{k}$, and this property was tested by numerical
calculation with a rectangular $\rho _{c}^{0}(\varepsilon ).$ The agreement
was perfect, indicating that one should be able to prove analytically this
property, but we have not made any attempt in such direction. A different
approach is to compare Eqs. (\ref{E4.10},\ref{E4.13}) with the corresponding
expressions that were obtained in the calculation of the CCHA \cite{FFM2}:
it was verified that the expressions for $n_{\sigma }(\bf{k})$ and for $%
D_{0\sigma }(\bf{k})$ are identical and therefore $n_{0}$, $n_{\sigma }$
and $D_{0\sigma }=n_{0}+n_{\sigma }$ are also equal in the two methods.
Since there is not an independent expression for $n_{\overline{\sigma }}$ in
the CCHA, the paramagnetic condition $n_{\overline{\sigma }}=n_{\sigma }$
was imposed \cite{FFM2} and completeness then reads as
\begin{equation}
D_{0\sigma }+n_{\sigma }=1\qquad ,  \label{E4.16}
\end{equation}
which was shown \cite{FFM2} to be satisfied for several $\rho
_{c}^{0}(\varepsilon )$ and $V(k)$. Since $D_{0\sigma }$ and $n_{\sigma }$
are identical in the two methods, this relation is then satisfied in the
CCHU when is true in the CCHA.

Equation (\ref{E4.12}) shows that when $U\rightarrow \infty $, completeness
(viz. Eq. (\ref{E3.7d})) is satisfied for any $\rho _{c}^{0}(\varepsilon )$
and $V(k)$ in the full space, and employing $n_{d}(\bf{k})=0$ (and
therefore $n_{d}=0$ ) there follows that

\begin{equation}
D_{0\sigma }+n_{\overline{\sigma }}=1\qquad .  \label{E4.17}
\end{equation}
Taking the difference of Eq. (\ref{E4.16}) and Eq. (\ref{E4.17}), it follows
that $n_{\overline{\sigma }}=n_{\sigma }$ in the CCHU and that these values
coincide with those calculated with the CCHA, at least for all the $\rho
_{c}^{0}(\varepsilon )$ and $V(k)$ for which Eq. (\ref{E1.2}) was
numerically verified for this approximation. This result shows that
employing the conjecture proposed in reference \cite{FFM2}, the results
obtained by the CCHA in the reduced space, that had the state $\mid
j,d\rangle $ projected out, are compatible with those calculated by the CCHU
in the full space. The occupation number of the two spin states is unique in
the CCHU, thus removing the difficulty presented by the CHA.

The inconsistency that is brought to light by the use of the CHU is the
difference between the values of $n_{\overline{\sigma }}$ and $n_{\sigma }$
obtained in that calculation. The dashed curves in figures \ref{F3} and \ref
{F4} plot the value of $n_{\sigma }$ calculated with the CHU, which we have
shown to be equal to the $n_{\sigma }$ obtained in the CHA, and it is set to
be equal to the $n_{\overline{\sigma }}$ in this last approximation because
that quantity can not be independently obtained in the reduced space. The
calculation in the full space makes possible to obtain an independent value
of $n_{\overline{\sigma }}$ in the CHU, which as discussed before is equal
to the unperturbed $n_{\sigma }^{0}$, that is shown in figures \ref{F3} and
\ref{F4} by the dotted curves. The different occupation of $n_{\overline{%
\sigma }}$ and $n_{\sigma }$ in the CHU is clearly shown at low $T$ in the
two figures, and these are the values that have to be employed to satisfy
completeness (viz. Eq. (\ref{E3.7d}) in the full space and Eq. (\ref{E1.2})
when $U\rightarrow \infty $). It is then clear that it is not possible to
satisfy Eq. (\ref{E1.2}) in this region when the relation $n_{\overline{%
\sigma }}=n_{\sigma }$ is forced by the CHA. The $n_{\sigma }$ in the CCHU
coincides with the $n_{\sigma }$ in the CCHA, and it is given by the full
line.

It is interesting to remark, that the terms with $A$ and $C$ in Eqs. (\ref
{E4.1} and \ref{E4.2}) are proportional to $\left\langle X_{00}\right\rangle
\ \left\langle X_{\sigma \sigma }\right\rangle /T=n_{\sigma }^{0}\
n_{0}^{0}\ /T$. As this contribution is proportional to $1/T$, it could be
dominant at very low $T$ when $0.1<\varepsilon _{f}/T<10$, but it is very
small outside this interval because either $\left\langle X_{00}\right\rangle
$ or$\ \left\langle X_{\sigma \sigma }\right\rangle $ is then much smaller
than $T$. The region inside this interval is the intermediate valence
region, and corresponds, at low $T$, to a crossover from the Kondo region to
a region of a thermally excited local moment. All the results are fairly
well behaved when $\varepsilon _{f}$ is not too small. The whole
contribution of Eq. (\ref{E4.1}) is equal to the difference between the
dashed curve ($n_{\sigma }$ for the CHU) and the full curve ($n_{\sigma }$
for the CCHU). This contribution is proportional to $n_{\sigma }^{0}$, and
the two curves coincide for very small $T$ in figure \ref{F3} while they
tend to different values in figure \ref{F4} because in this case the $%
n_{\sigma }^{0}$ does not vanish for $T\rightarrow 0$.

The value of $n_{\sigma }$ as a function of $\mu $ for $T=0.1$ is plotted in
figure \ref{F6} for the same approximations used in the previous figures,
showing a monotonic increase with $\mu $ as it could be expected.

\begin{figure}[H]
\caption[Fig.6]{The occupation number $n_{\sigma }$ is plotted as a function
of $\mu $ for $T=0.1$ while all the remaining parameters are the same used
in figure~\ref{F2}. All the curves are monotonically increasing functions of
$\mu $. }
\label{F6}
\end{figure}

\vspace*{1pt}

\section{Concluding Remarks}
\vspace*{-0.5pt} \noindent
\noindent
The Anderson lattice is a model that gives a schematic description of many
real systems with strongly correlated electrons, and it seems important to
understand better the different methods employed to calculate its
properties. In this work we consider the cumulant expansion, a perturbative
method that has many desirable properties, and we employ the Hubbard
operators, that make it easy to project the space of the system into a
subspace of states of interest. The main motivation of using the Hubbard
operators is that they are very well suited to reduce the space of states of
the system by eliminating those that are no relevant to the properties
considered.

There are several perturbative expansions that use this operators,
and here we shall briefly compare three of them: \cite
{Hubbard5,IzyumovLSBC,KeiterM}. All these expansions employ the
Matsubara technique\cite{AbrikosovGD} with imaginary times, and
first expand the grand partition function or the Green's functions
as a series of statistical averages of time ordered products of
operators that contain successive orders of the interaction
Hamiltonian. The technique developed by Keiter, Kimball and Grewe
and reviewed in reference \cite{KeiterM}, expresses these averages
in terms of a diagramatic expansion that leads to a scheme of the
Brillouin-Wigner type\cite {GreweK,RamakrishnanS}. This method has
been already compared with our treatment in more detail \cite{FFM}
, and shall not be further discussed here.

We shall consider now the method in \cite{IzyumovLSBC}, and show
that it employs a type of cumulants different from the ones used in
our treatment.The initial averages are reduced in
\cite{IzyumovLSBC} by a technique already used by
Gaudin\cite{Gaudin} to prove the standard Wick 's theorem, and
later applied to averages of spin operators\cite{YangW} and to
Hubbard operators of the Fermi type\cite{Hewson,FFM2}. This
technique takes advantage of the fact that the commutator of two
Hubbard operators is a linear combination of the same type of
operators, making it possible to reduce an average with $n$ Hubbard
operators into a linear combination of averages with only $n-1$ of
them. In \cite{IzyumovLSBC} they proceed with this reduction until
only averages of diagonal Hubbard operators (that are therefore of
the Bose type) remain, and these averages are then expressed in
terms of cumulants that can be obtained from an explicitly given
generating function. A diagramatic expansion that employs these
cumulants at single sites is then obtained, and a renormalization
of vertices can be also performed\cite {BartkowiakC,BartkowiakC2}.

Our expansion, on the other hand, closely follows the original
derivation employed for the Ising
model\cite{Wortis,Englert,BlochL}, and the initial averages are
expressed directly in terms of cumulants, employing a general
property that was extended to include cumulants with Hubbard
operators of the Fermi type( cf. Theorem 3.1 in reference
\cite{FFM}). A diagramatic expansion different from that in
\cite{IzyumovLSBC} is then obtained, and the same reduction
method\cite{Hewson,FFM2} is employed to calculate the expressions
of the cumulants.

 Although the two methods are related, the one described in
\cite{IzyumovLSBC} is different from ours, and their cumulants
correspond to different operators. The cumulants in
\cite{IzyumovLSBC} only contain ``diagonal operators'' that are of
the Bose type, while the Hubbard operators that appear in the
cumulants of our expansion are of the Fermi type. In particular,
the cumulants that are of order $n$ in \cite{IzyumovLSBC} are of
order $2n$ in our treatment. As discussed in the Appendix for the
CHA, the relations that are related by Englert to the classical
Ward identities take in our case a rather different physical
meaning than the one corresponding to the Ising problem, because in
that problem the momentum operators are linear combination of
Hubbard operators of the Bose type, while they are of the Fermi
type in our case.

One difficulty in the cumulant expansion we employ, is that the approximate
GF obtained do not usually satisfy completeness (i.e. conservation of
probability). In the present work we consider the cumulant expansion of the
Anderson lattice, and employ the Hubbard operators to project, in the limit
of $U\rightarrow \infty $, the unoccupied state with two electrons out of
the space of local electrons. It has been shown \cite{FFM2} that the ``chain
approximation'' (CHA), which corresponds to the most general family of
diagrams with only second order cumulants, does not satisfy completeness in
the reduced space. It was useful to analyze the problem by keeping the
doubly occupied state for finite $U$ within this approximation, and then
taking $U\rightarrow \infty $. It was shown that completeness is satisfied
when this procedure is used, but to obtain this result it is necessary to
use different values of $n_{\sigma }$ and $n_{\overline{\sigma }}$, that are
obtained respectively employing Eq. (\ref{E2.5}) and Eq. (\ref{E2.7}) with
the two GFs $G_{0\sigma ,0\sigma }$ and $G_{\overline{\sigma }d,\overline{%
\sigma }d}$ associated with the creation or destruction of the same spin $%
\sigma $ (cf. Eq. (\ref{E3.7e})). To exemplify this situation employing
either of figures \ref{F3} or \ref{F4}, note that to satisfy completeness it
is necessary to use the dotted curve for one spin and the dashed curve for
the opposite spin. When the GF is calculated in the reduced space (without
the state $\mid d\rangle $) there is no $G_{\overline{\sigma }d,\overline{%
\sigma }d}$, and the paramagnetic condition $n_{\sigma }=n_{\overline{\sigma
}}$ has to be forced on the system. The dashed line has then to be used for
the occupation of both spin components, and completeness is therefore not
satisfied when the dashed curve is different from the dotted one. One
important consequence of our derivation is that although the calculations
employing the full space do satisfy completeness, they might hide the use of
inconsistent values of $n_{\sigma }$ and $n_{\overline{\sigma }}$.

A conjecture on how to obtain families of diagrams that satisfy completeness
has been proposed \cite{FFM2}, and it was verified for some families of
diagrams that might contain any number of fourth order cumulants \cite
{Figueira}. The ``complete chain approximation'' satisfies completeness and
was obtained by adding diagrams to the CHA according to the proposed
conjecture. It seemed then relevant to analyze the completeness problem in
the CCHA by considering again the full space for finite $U$ and then taking $%
U\rightarrow \infty $. It was found that the $n_{\sigma }$ and $n_{\overline{%
\sigma }}$ obtained when employing the two GFs $G_{0\sigma ,0\sigma }$ and $%
G_{\overline{\sigma }d,\overline{\sigma }d}$ satisfy $n_{\sigma }=$ $n_{%
\overline{\sigma }}$ and are also equal to the $n_{\sigma }$ that was
calculated with the CCHA, which only employs $G_{0\sigma ,0\sigma }$. We
conclude that the inconsistency pointed out for the CHA disappears for the
CCHA. Moreover, we could find the occupation number $n_{a}(\bf{k})$
associated to the wavevector $\bf{k}$ and the state $\mid a\rangle $ for
$a=0,\sigma ,\overline{\sigma }$,(cf. Section~\ref{SD.2}) because it is
possible to calculate those values analytically. Employing $G_{0\sigma
,0\sigma }$ we have also found that the $n_{\sigma }(\bf{k})$ and $n_{0}(%
\bf{k})$, calculated in CCHU are identical to those calculated employing
the CCHA.

The interest in the use of the Hubbard operators is that given a model with
local states, it is a fairly simple procedure to reduce the corresponding
space of states to a subspace that only contains the few states that are
considered more relevant; e.g.: to study the Jahn-Teller effect of $Co^{2+}$
in $MgO$, the 120 states of the $(3d)^{7}$ ground configuration of $Co^{2+}$
have been projected into the 12 ground states that are split by the Coulomb
interaction and by the crystal fields \cite{Foglio,Luciano}. The question
immediately arises as to what contributions are lost by the use of this
method. The technique employed in the CCHU, of comparing the results in the
reduced space with those obtained by calculating the properties in the full
space and then making the energy of the states that would be removed go to
infinity, seems a good test. Although our conclusions strictly apply only to
the problem considered, the results summarized in the previous paragraphs
show that one can safely use the reduced space when the occupation of the
removed states goes to zero.

\nonumsection{Acknowledgements}
\noindent
The authors are grateful to Prof. Roberto Luzzi for critical comments. They
would like to acknowledge financial support from the following agencies:
CAPES-PICD (MSF), FAPESP and CNPq (MEF). This work was done (in part) in the
frame of Associate Membership Programme of the International Centre for
Theoretical Physics, Trieste ITALY (MEF).

\nonumsection{References}

\appendix
\noindent

Properties of the Chain Approximation

Although the main purpose of the present paper is not to discuss the CHA,
and much less to propose it for the calculation of all the Anderson lattice
properties, this approximation has some interesting characteristics that
seem worth while presenting in a rather summary way. Within the context of
the cumulant expansion of the Anderson lattice model, the CHA was studied by
Hewson\cite{Hewson} and its was recently analyzed in more detail\cite
{FFM,FFM2}.

From the point of view of the high density $1/z$ classification of
diagrams \cite{Englert,IzyumovLSBC,Stinchcombe,CottamS} the CHA is
of zeroth order. An interesting feature is that the CHA is a $\Phi
-$derivable approximation\cite{FFM2}, generated by the same type of
skeleton that gives the molecular field approximation in the
Ising\cite{Wortis} and in the quantum Heisenberg model%
\cite{Wortis,StinchcombeHEB}. Although the present treatment of
the Anderson lattice closely follows the cumulant expansion for the
Ising model \cite{Englert}, there is an essential difference with
the treatment employed for the magnetic systems, namely that the
external fields we employ to generate the
cumulants\cite{Hubbard5,FFM,Metzner} require the introduction of
the Grassman fields $\xi $, that have no physical meaning and
should be set to zero at the end of the calculation. The operators
associated to those fields are not magnetic moments, which do not
change the number of electrons, but they create or destroy an
electron, so that their averages have to be zero when all the $\xi
=0$. As a consequence, one does not obtains a self consistent
relation equivalent to the Weiss molecular field equation, and for
the same reason many of the relations established\cite
{Wortis,Englert,BlochL} for the Ising model change its meaning in
the present context. Nevertheless, the CHA retains some of the
characteristics of a mean field approximation\cite{FFM2}: the
correlation between the electrons, that in the limit $U=\infty $
would forbid the occupancy of a local electron with spin $\sigma $
at any given site that already has a local electron with spin
$-\sigma $, is replaced by an average reduction of the
hybridization constant $V^{2}$ by a factor $D_{\sigma
}=1-n_{\overline{%
\sigma }}$, equal for all the sites.

When the general formalism of the diagramatic expansion is replaced
by an approximate evaluation, many of the general properties might
be lost in the procedure, and one could make a list of properties
that would be desirable to maintain \cite{Bartkowiaktes}. It seems
then interesting to analyze the CHA from that perspective, and in
particular by considering that it is a $\Phi -$ derivable
approximation.

The study of the $\Phi -$derivable approximations\cite{BlochL,Baym}
in the cumulant expansions for the Ising model has been reviewed by
Wortis\cite {Wortis}, and the corresponding treatment for the
Hubbard model\cite{Metzner} and for the Anderson lattice\cite{FFM2}
have been also recently discussed, so we shall use the concepts and
notations presented in those references to avoid repetition. The
basic idea is to replace the many unrenormalized diagrams of the
cumulant expansion by fewer skeleton diagrams with renormalized
vertices $M_{n}$, and employ a functional $\Phi $ of the $M_{n}$
that corresponds to the complete family of skeletons, to obtain the
grand canonical potential $\Omega =T\ \ln \cal{Z}$, where the
$\cal{Z }$ is the grand partition function. In the $\Phi
-$derivable approximation only a subfamily of all the possible
skeletons is considered, and three relations that are valid in the
exact expansion are employed to define completely the approximate
method\cite{Wortis}. These three conditions relate the $\ln
\cal{Z},$ $M_{n}$ and the ``self-fields'' $S_{n}$ \cite{Wortis,Metzner},
and correspond to the Eqs.(3.1,3.2,3.5) given in reference \cite{FFM2} for
the exact expansion. One would then expect that as the correlation functions
in a $\Phi -$derivable approximation are obtained from ``the same underlying
free energy'', they ``will have singular behavior at the same points and
obey certain self-consistency conditions''\cite{Wortis}.

Variational properties for the free energy that are valid for the Ising model%
\cite{Englert,BlochL} and for the Heisenberg model\cite{Stinchcombe}, can be
easily extended to the $\Phi -$derivable approximations of the Anderson
lattice by making the appropriate changes to the notation\cite{FFM2}. These
properties guarantee the relation (cf. Eq. (A.28) in \cite{FFM} and Eq.
(3.7) in \cite{FFM2})
\begin{equation}
\frac{\delta \ln [\cal{Z}(\beta ,\xi )]}{\delta \xi (\ell )}\equiv
\left\langle \widehat{Y}(\ell )\right\rangle _{c}^{V,\xi }=M_{1}\left( \ell
;\xi \right)  \label{A.1}
\end{equation}
which corresponds to the Eq. (4.17) in \cite{Englert}. Although this
relation equates the magnetization to the renormalized cumulant $M_{1}$ in
the Ising model, it only gives the trivial identity $0=0$ when $\xi =0$ in
the present cumulant expansion. Nevertheless, for $\xi \neq 0$ one can employ

\begin{equation}
\frac{\delta ^{2}\ln [\cal{Z}(\beta ,\xi )]}{\delta \xi (\ell )\ \delta
\xi (\ell ^{\prime })}\equiv \left\langle \widehat{Y}(\ell )\ \widehat{Y}%
(\ell ^{\prime })\right\rangle _{c}^{V,\xi }=\frac{\delta M_{1}\left( \ell
;\xi \right) }{\ \delta \xi (\ell ^{\prime })}\qquad ,  \label{A.2}
\end{equation}
and, following the Appendix A in reference \cite{FFM2}, obtain the Eq. (\ref
{E3.1}) employed in the present work in the CHA for $\xi =0$, or the Eqs. (%
\ref{E3.4}-\ref{E3.7}) for the CHU. One could also calculate $\left\langle
\widehat{Y}(\ell )\ \widehat{Y}(\ell ^{\prime })\right\rangle _{c}^{V,\xi }$
directly from the diagramatic expansion, and the two results should be
equal; Englert has shown that in the Ising model this equality requires a
relation between the vertex and the bond renormalization ( the Eq.(5.22) in
reference \cite{Englert}), which is very similar to the ``Ward identities''
of quantum electrodynamics. These identities have been also discussed by
Stinchcombe\cite{Stinchcombe} for the Heisenberg model. It is a simple
matter to show that both the CHA and CHU satisfy the Ward identity in the
form stated by Englert, because the only non zero self-fields in the vertex
renormalization are the $S_{1}$, and the only diagrams that contribute to
the GF $\left\langle \widehat{Y}(\ell )\ \widehat{Y}(\ell ^{\prime
})\right\rangle _{c}^{V,\xi =0}$ are the simple chains in figure \ref{F1},
so that the renormalized bond is the same as the bare bond. Another way to
prove this relation is to notice that the GF in the diagramatic calculation%
\cite{FFM} are equal to those obtained from the $\Phi -$derivable
approximation employing the simple skeleton of Figure 3a in reference \cite
{FFM2}, both for the CHA and for the CHU.

Another interesting aspect of the CHA is the calculation of the $\ln
\cal{Z}$, that would make possible to derive all the thermodynamic
properties of the system. By the very simplicity of the skeleton that gives
the CHA and the CHU, the only non zero self-fields in this approximation are
the $S_{1}$\cite{FFM2}, and these are also zero in the physical region when
all $\xi =0$, so that $\ln \cal{Z}=\ln $${\cal{Z}}_{0}$, i.e. we obtain
the same grand partition function ${\cal{Z}}_{0}$ of the unperturbed
system. All the thermodynamic relations are then automatically satisfied,
but at the price of not giving any new information. To obtain a better $\ln
\cal{Z}$ one can use the GF in the CHA, and the coupling constant
integration method has been employed to this purpose in reference \cite
{Physica}. Another possible method is to integrate
\begin{equation}
\left( \frac{\partial F}{\partial N_{t}}\right) _{T,V}=\mu \qquad ,
\label{A.3}
\end{equation}
where $N_{t}$ is the total number of electrons in the system. After solving
a puzzling paradox that appears in a parameter region where three solutions
are possible for a given $N_{t}$, one can compare the free energies $F$
obtained by these two methods of calculation. In the reference mentioned,
their values are plotted for the atomic limit and compared with the available
exact solution, showing that the two methods give fairly close values of $F$
that compare reasonably well with the exact curve.

Another very important property of the approximation is the spectral density
of the one-electron GF, which is plotted in Figure \ref{F2} for the f
electrons, in the CHA. This approximation gives the gross behaviour of the
spectral density, but it misses the presence of the Kondo peak, that should
appear close to the Fermi surface at low $T$ when the system parameters are
in the Kondo region. This peak is what gives the characteristically large
susceptibility and specific heat of the heavy fermions at low $T$, and its
absence from the CHA and CHU is to be expected because in this approximation
there are no spin flips, that are essential for the existence of the Kondo
effect.

To improve on the molecular field approximation one should add the
rings\cite {Wortis,Englert}, that are the next order diagrams in
the $1/z$ expansion. A family with infinite fourth order cumulants
has been considered\cite{FFM}, and the corresponding spectral
density shows some differences with that of the CHA, but it does
not show a Kondo peak either, as can be seen in the figure 5 of
reference \cite{Foglio}. The same problem is therefore present in
both the CCHA and the CCHU. One should also mention that employing
a similar expansion, the magnetization of the Hubbard model has
been calculated by including all the cumulants\cite{BartkowiakC},
but neglecting the spin fluctuations in their calculation: this
type of technique would not be then able to recover the Kondo peak
in the Anderson lattice. By a similar approximation the exact
solution of the Falicov Kimball model has been
obtained\cite{CracoG2}, and the spin fluctuations of the Hubbard
model have been approximately included in the same work.

The absence of the Kondo peak from the approximate GF that have
employed is still under study, and we have been able to find an
approximation closely related to the cumulant expansion that shows
a structure near the Fermi surface that decreases in intensity when
$T$ increases\cite{Foglio}, and could therefore be identified as
the Kondo peak.

The present summary shows that the CHA and CHU have very special
properties within the cumulant expansion, and the absence of
completeness in the CHA seemed particularly intriguing, thus giving
the motivation for the present work, that serves to clarify this
behavior.

The CCHA is not $\Phi -$derivable, but is satisfies completeness and carries
within its structure the elimination of some inconsistencies of the CHA that
are only revealed through the study of the CCHU. It seems clear that many
valuable approximations would not satisfy all the properties of the exact
solution that one would like to maintain, but we believe that they are
nevertheless worth its study and use, provided that their shortcomings are
kept in mind.

\newpage


\noindent
\begin{thebibliography}{99}

\bibitem{Hubbard5}  J. Hubbard, {\it Proc. R. Soc. London, Ser. }\rm{A%
} {\bf{296}}, 82 (1966).

\bibitem{Foglio}  M. E. Foglio,{\it \ Brazilian Journal of Physics}
accepted for publication {\bf{8}}, (1997)

\bibitem{CracoG}  L. Craco and M. A. Gusm\~{a}o, {\it Phys. Rev. }\rm{%
B} {\bf{52}}, 17135 (1995).

\bibitem{Hewson}  A. C. Hewson, {\it J. Phys. C: Solid State Phys.}
{\bf{10}}, 4973 (1977).

\bibitem{FFM}  M. S. Figueira, M. E. Foglio and G. G. Martinez, {\it %
Phys. Rev. }\rm{B} {\bf{50}}, 17933 (1994).

\bibitem{Reviews}  P.A. Lee, T.M. Rice, J.W. Serene, L. J. Sham and J. W.
Wilkins, {\it Comments Cond. Mat. Phys. }{\bf{12}}, 99 (1986).

P. Fulde, {\it Solid State Physics.} {\bf{41}}, 1 (1988).

P. Schlottmann, {\it Phys.Rep.} {\bf{181}}, 1 (1989).

A. C. Hewson,{\it The Kondo problem to Heavy Fermions} (Cambridge U.P.
Cambridge 1993).

A. Georges, G. Kotliar, W. Krauth and M. J. Rozenberg. {\it Rev. Mod.
Phys. }{\bf{68}}. 13 (1996).

\bibitem{Wortis}  M. Wortis, in {\it Phase Transitions and Critical
Phenomena}, edited by C. Domb and M. S. Green (Academic, London, 1974), Vol.
{\bf{3}}. pg. 113.

\bibitem{Englert}  F. Englert, {\it Phys. Rev. }{\bf{129}}, 567 (1963).

\bibitem{StinchcombeHEB}  R. B. Stinchcombe, G. Horwitz, F. Englert and R.
Brout, {\it Phys. Rev. }{\bf{130}}, 155 (1963).

\bibitem{Hubbard123}  J. Hubbard, {\it Proc. R. Soc. London, Ser. }%
\rm{A} {\bf{276}}, 238 (1964); {\it ibid. }\rm{A} {\bf{277}},
237 (1964); {\it ibid. }\rm{A} {\bf{281}}, 401 (1964).(These are
the first three papers of a series of six).

\bibitem{Fulde}  P. Fulde, {\it Electron correlations in Molecules and
Solids} (Springer Verlag. 1995).

K. W. Becker and P. Fulde, {\it Z. Phys. B.} {\bf{72}}, 423 (1988).

K. W. Becker and P. Fulde, {\it J. Chem. Phys.} {\bf{91}}, 4223 (1988).

K. W. Becker and W. Brenig, {\it Z. Phys. B.} {\bf{79}}, 195 (1990).

\bibitem{IzyumovLSBC}  Yu. A. Izyumov, B. M. Leffulov, E. V. Shipitsyn, M.
Bartkowiak and K. A. Chao, {\it Phys. Rev. }\rm{B} {\bf{46}}, 15697
(1992).

\bibitem{BartkowiakC}  M. Bartkowiak and K. A. Chao, {\it Phys. Rev. }%
\rm{B} {\bf{47}}, 1616 (1993).

\bibitem{Metzner}  W. Metzner, {\it Phys. Rev. }\rm{B} {\bf{43}},
8549 (1991).

\bibitem{Enrique}  E. V. Anda, {\it J. Phys. C: Solid State Phys.}
{\bf{14}}, L1037 (1981).

\bibitem{FFM2}  M. S. Figueira and M. E. Foglio, {\it J.\ Phys.: Condens.
Matter } {\bf{8}}, 5017 (1996).

\bibitem{Negele}  J. W. Negele and H. Orland, {\it Quantum Many-Particle
Systems} (Addison-Wesley, New York, 1988), Chap. 2.

\bibitem{YangW}  D. H. Y. Yang and Y. L. Wang, {\it Phys. Rev.} B {\bf{%
10}}, 4714 1974.

\bibitem{Physica}  M. S. Figueira and M. E. Foglio, {\it Physica }A
{\bf{208}}, 279 (1994).

\bibitem{YangW75}  D. H. Y. Yang and Y. L. Wang, {\it Phys. Rev.} B
{\bf{12}}, 1057 1975.

\bibitem{BartkowiakC2}  M. Bartkowiak and K. A. Chao, {\it Phys. Rev. }%
\rm{B} {\bf{47}}, 4193 (1993).

\bibitem{Stinchcombe}  R. B. Stinchcombe, {\it \ J. Phys. C: Solid State
Phys.} {\bf{3}}, 2266 (1970).

\bibitem{CottamS} M. G. Cottam and R. B. Stinchcombe,
{\it \ J. Phys. C: Solid State Phys.} {\bf{3}}, 2283 (1970).

\bibitem{KeiterM}  H. Keiter and G. Morandi {\it Phys. Rep.}
{\bf{109}}, 227 (1981).

\bibitem{AbrikosovGD}  A. A. Abrikosov, L. P. Gorkov and I. E.
Szyaloshinski, {\it Methods of Quantum Field Theory in Statistical Physics%
} (Dover, 1975).

\bibitem{GreweK}  N. Grewe and H. Keiter, {\it Phys. Rev. }\rm{B}
{\bf{24}}, 4420 (1981).

\bibitem{RamakrishnanS}  T. V. Ramakrishnan and K. Sur, {\it Phys. Rev. }%
\rm{B} {\bf{26}}, 1798 (1982).

\bibitem{Gaudin}  M. Gaudin, {\it J. Nucl. Phys.} {\bf{15}}, 89 (1960).

\bibitem{BlochL}  C. Bloch and J. S. Langer, {\it \ J. Math. Phys.}
{\bf{6}}, 554 (1965).

\bibitem{Figueira}  M. S. Figueira, Ph.D. thesis, Universidade Estadual de
Campinas, SP, Brazil, 1994.

\bibitem{Luciano}  L. T. Peixoto and M. E. Foglio,{\it \ Phys. Rev.}
\rm{B }{\bf{32}}, 2596 (1985)

\bibitem{Bartkowiaktes}  M. Bartkowiak - ``High density expansion for narrow
band systems '' - Ph. D. Thesis - Adam Mickiewicz University - Poznan,
Poland, 1988.

\bibitem{Baym}  G. Baym, {\it \ Phys. Rev.} {\bf{127}}, 1391 (1962).

\bibitem{CracoG2}  L. Craco and M. A. Gusm\~{a}o, {\it Phys. Rev. }%
\rm{B} {\bf{54}}, 1629 (1996).

\end{thebibliography}
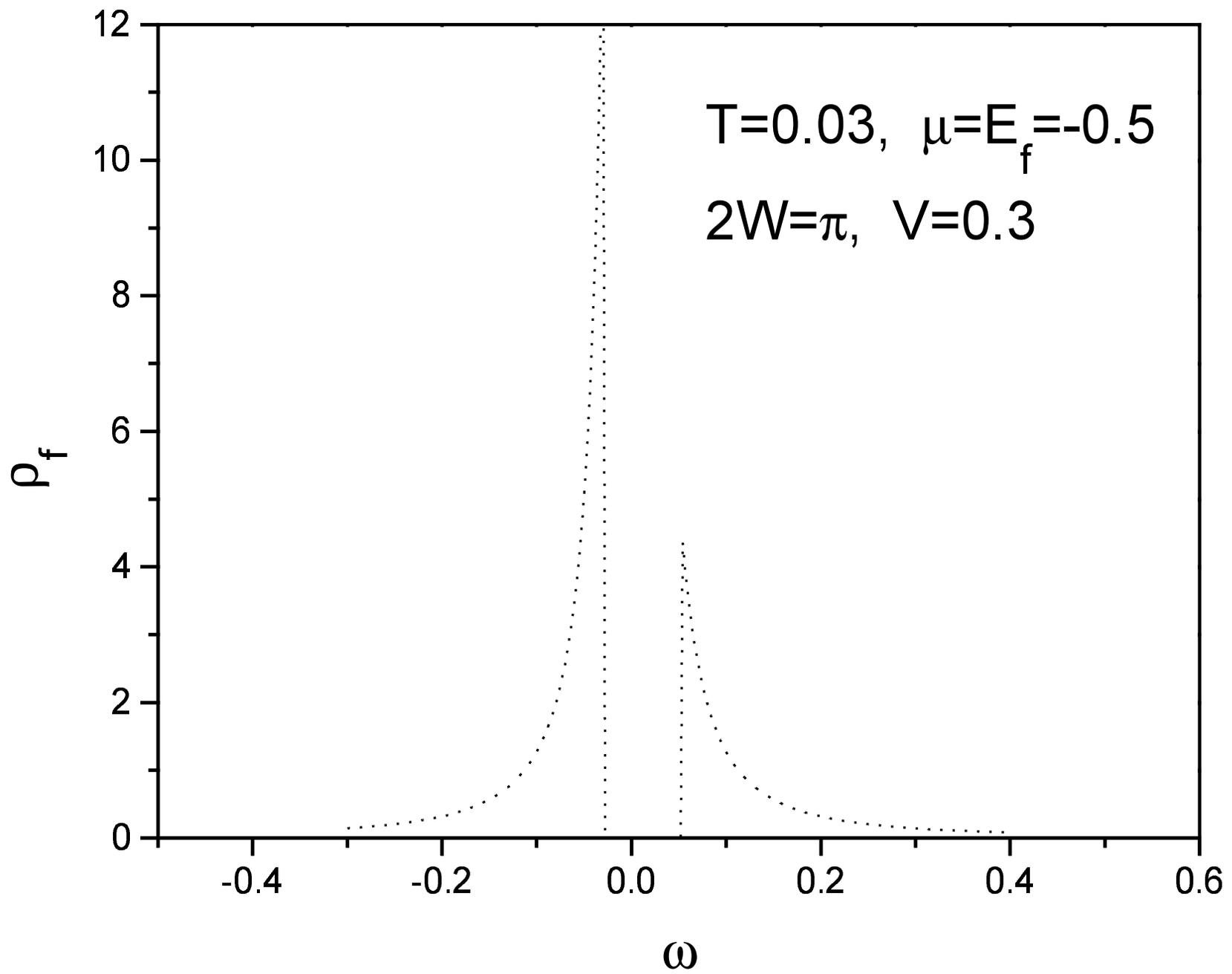
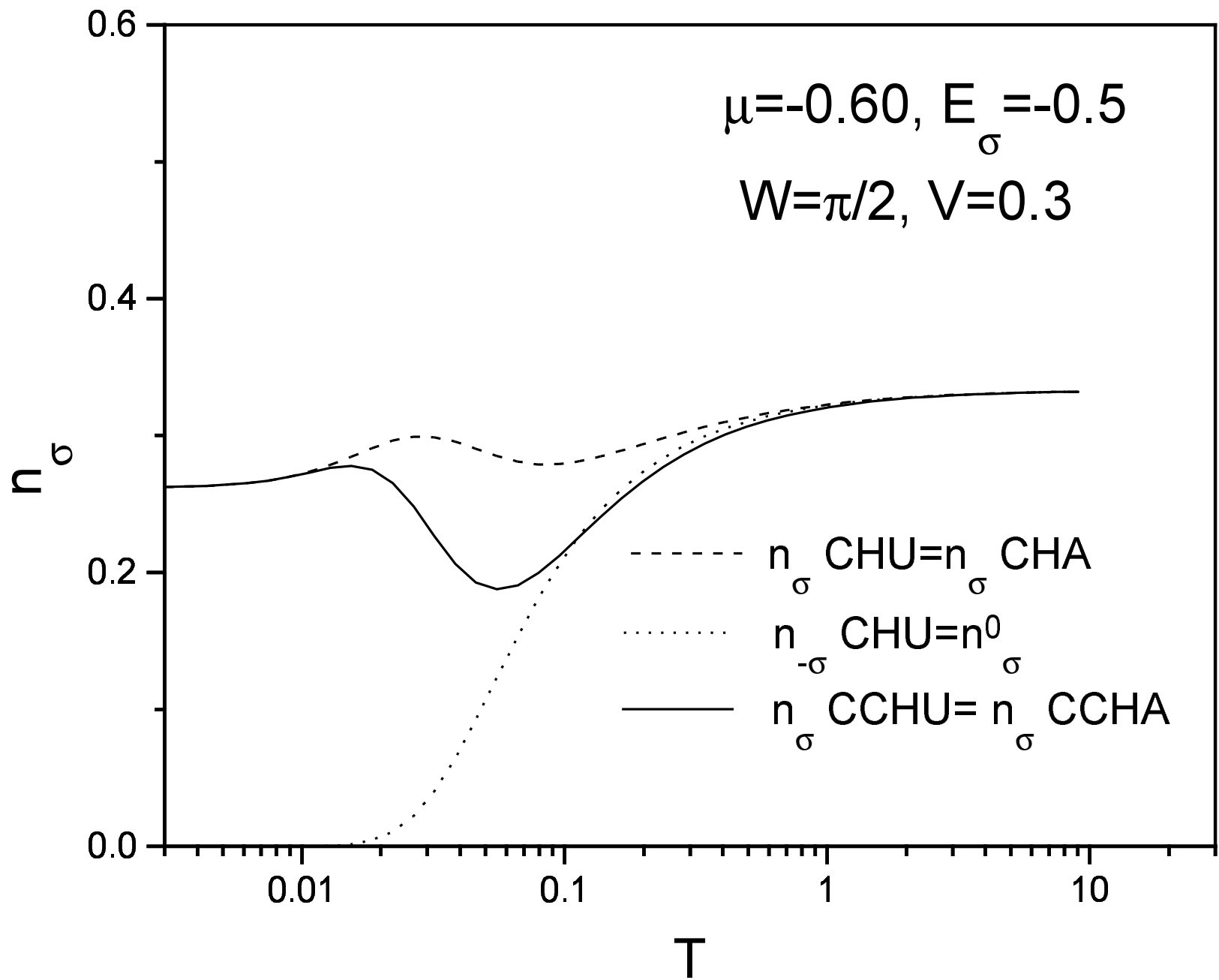
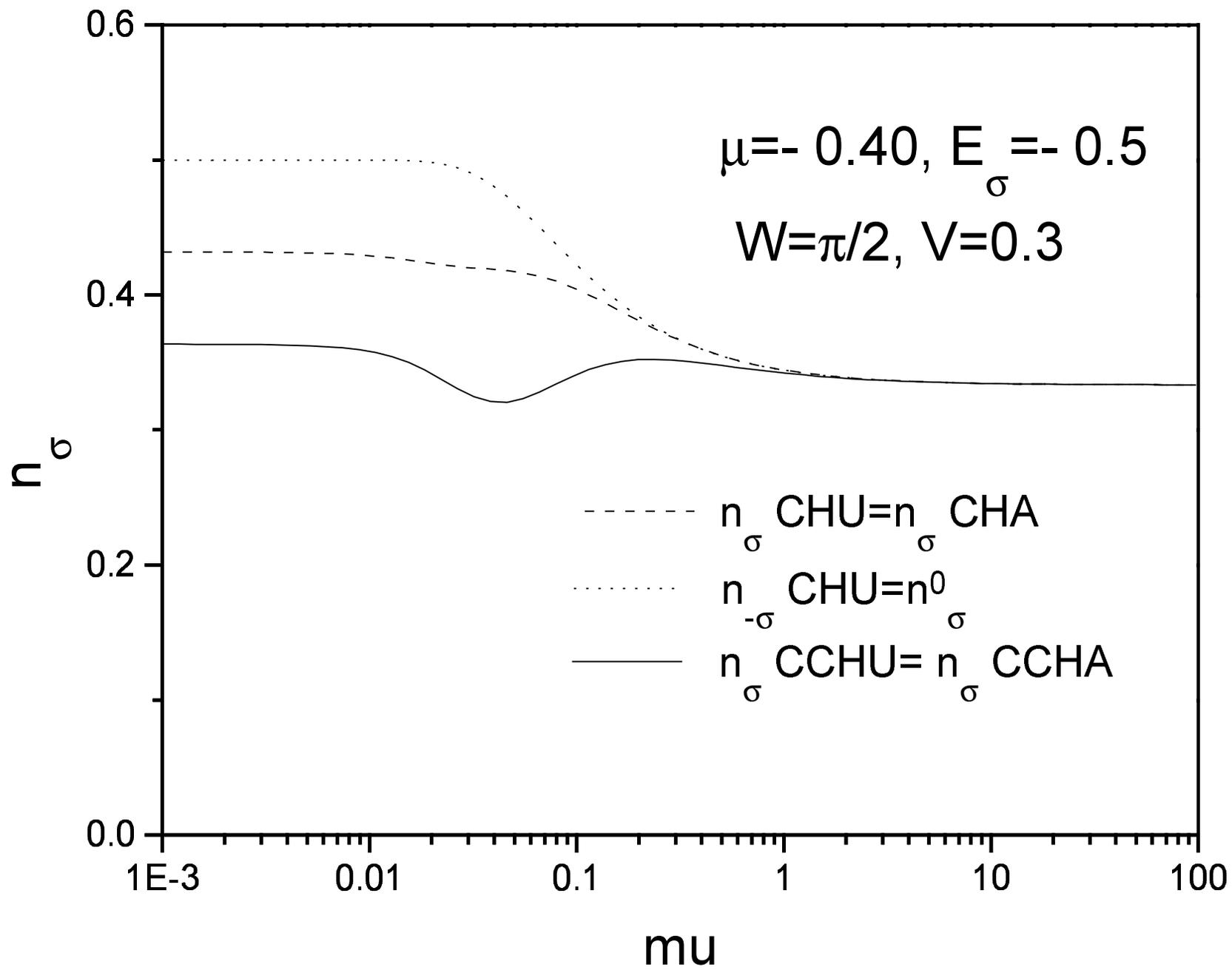
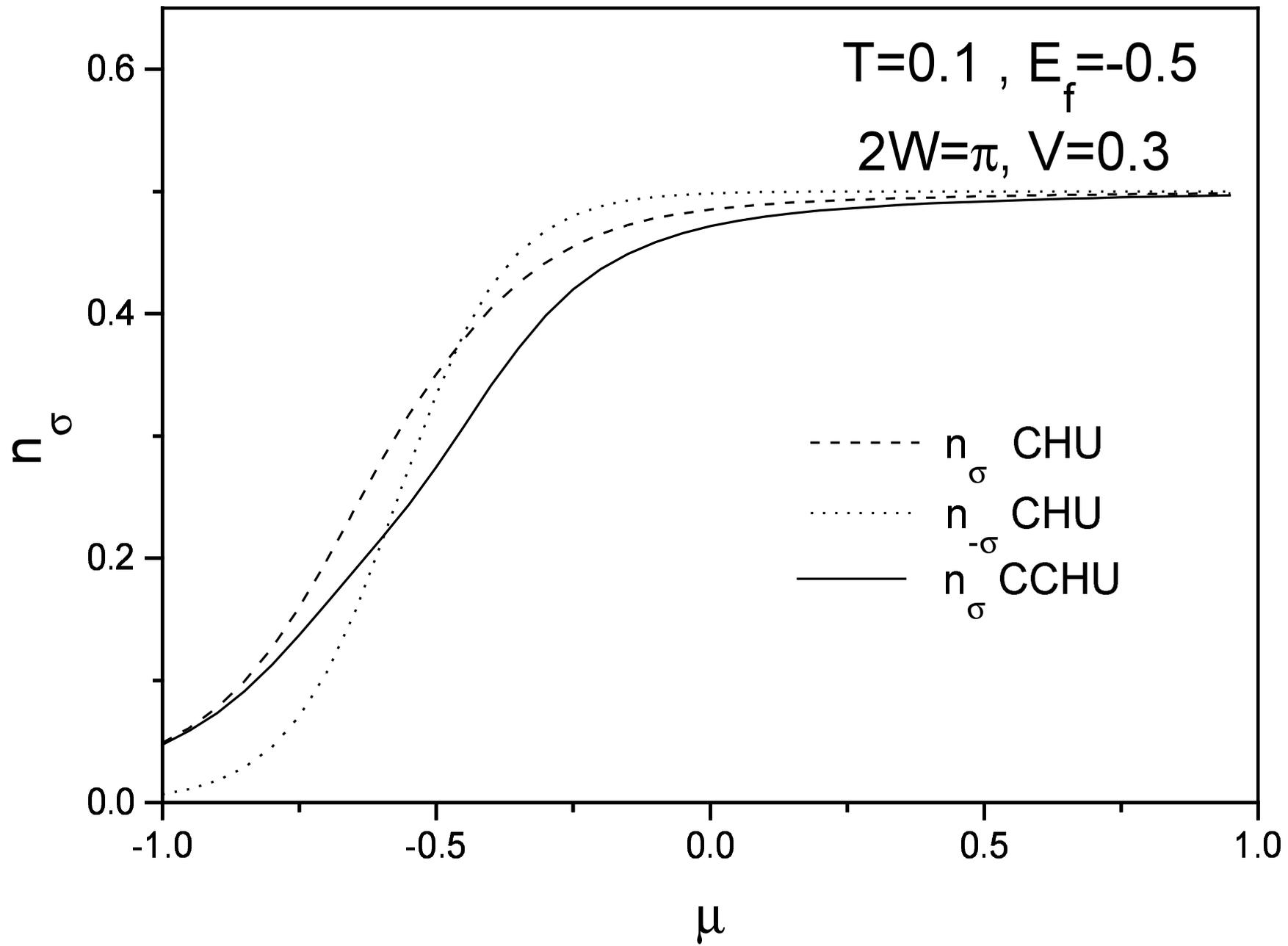
\end{document}